\documentclass[3p,times]{elsarticle}
\usepackage{graphicx}
\usepackage{sidecap}
\usepackage{amssymb,amsmath,amsthm,amsbsy}
\usepackage{calligra,calrsfs,mathrsfs}
\usepackage{multirow}
\usepackage{ulem,cancel}
\usepackage[usenames,dvipsnames]{color}
\usepackage{copyrightbox}
\usepackage{cases}
\usepackage{pgfplots}
\usepackage{subcaption}
\usepackage{tikz} 
\usepackage{tkz-euclide}
\usepackage{tikz-dimline}
\usetikzlibrary[patterns]
\usepackage{hyperref}
\usepackage{algorithm}
\usepackage{algpseudocode}
\usepackage{dsfont}
\usepackage{caption}
\usepackage{epstopdf}
\usepackage{epsfig}
\usepackage{natbib}
\usepackage{grffile}
\usepackage{comment}
\usepackage{enumerate}
\usepackage{array}
\newcolumntype{P}[1]{>{\centering\arraybackslash}p{#1}}
\usepackage{float}
\usepackage[T1]{fontenc}
\usepackage[latin9]{inputenc}
\usepackage{geometry}
\usepgfplotslibrary{fillbetween}
\usetikzlibrary{patterns}
\usepackage{color}
\usepackage{stmaryrd}
\usepackage{setspace}
\usepackage{amsthm}
\usepackage{textcomp}
\usepackage{hyperref}
\usepackage{soul,xcolor}
\usepackage[inline]{enumitem}
\usepackage{textcomp}
\usepackage{adjustbox}
\usepackage{todonotes}
\usepackage{empheq}
\graphicspath{{./}}

\allowdisplaybreaks

\newcommand{\ifcomment}{\iffalse}

\newcommand{\algai}{\hspace{\algorithmicindent}}

\newdefinition{rem}{Remark}

\biboptions{sort&compress}

\journal{Not Yet}

\begin{document}

\begin{frontmatter}

\title{Modeling Ostwald Ripening Dynamics in Porous Microstructures}
\author[psu]{Md Zahidul Islam Laku}
\ead{mxl6146@psu.edu}
\author[mcmaster]{Mohammad Salehpour}
\ead{salehpom@mcmaster.ca}
\author[mcmaster]{Tian Lan}
\ead{lant20@mcmaster.ca}
\author[mcmaster]{Benzhong Zhao}
\ead{robinzhao@mcmaster.ca}
\author[psu]{Yashar Mehmani}
\ead{yzm5192@psu.edu}
\address[psu]{Department of Energy and Mineral Engineering, The Pennsylvania State University, University Park, USA}
\address[mcmaster]{Department of Civil Engineering, McMaster University, Hamilton, ON, Canada}
\cortext[ca]{Corresponding author: Yashar Mehmani. Email: yzm5192@psu.edu}

\begin{abstract}
Partially miscible ganglia trapped in a porous medium evolve through Ostwald ripening, driven by differences in interfacial curvature. In practice, ganglia can span multiple pores and undergo discrete capillary events --- invasion, snap-off, retraction, fragmentation, coalescence, and dislocation --- that alter their topology and induce local flow. Existing pore-network models (PNMs) for ripening are limited to single-pore ganglia, assume idealized pore shapes, and operate under quasi-static conditions that preclude flow. We present an image-based pore-network model (iPNM) that removes these limitations. Unlike existing PNMs, iPNM requires no idealization of pore shapes, as the effect on capillarity is encoded locally in curvature--saturation curves computed via the pore-morphology method. iPNM couples two-phase flow, solute transport, and Ostwald ripening within a unified framework. We first verify iPNM against a prior quasi-static PNM, then validate it against recent high-resolution microfluidic experiments of hydrogen ripening in a sandstone-patterned micromodel over 15--24 days at 40$^\circ$C and 80$^\circ$C. Good agreement is obtained without adjustable parameters. Comparison with a continuum model shows that while macroscopic saturation is captured by both approaches, iPNM uniquely resolves population statistics, individual ganglion curvatures, and pre-equilibrium ripening dynamics within a representative elementary volume. Its computational efficiency over direct numerical simulation makes it suitable for guiding the development of improved theories of ripening in confined geometries.
\end{abstract}

\begin{keyword}
Porous media, Ostwald ripening, Pore-network model, Hydrogen storage, Ganglia, Pore morphology
\end{keyword}

\end{frontmatter}

\section{Introduction}
\label{sec:intro}

Partially miscible ganglia trapped within a liquid-saturated porous medium arise in several energy and environmental applications. In underground hydrogen storage (UHS), hydrogen is cyclically injected into and withdrawn from an aquifer to buffer the intermittency of renewable energy supply.
Each injection-withdrawal cycle leaves behind a population of disconnected ganglia whose long-term fate determines storage efficiency \cite{zivar2021uhs}.
In geological carbon storage, residual trapping of CO$_2$ ganglia is a primary mechanism to render the injected CO$_2$ hydrodynamically immobile \cite{bachu2008co2,iglauer2011gang}. In polymer electrolyte membrane fuel cells and water electrolyzers, oxygen and hydrogen produced at catalytic sites must traverse porous diffusion layers, and managing trapped gas is central to device performance \cite{andersson2016FC}. In all these settings, trapped ganglia exchange mass and evolve through a process known as Ostwald ripening \cite{ostwald1897}.

Ripening is driven by differences in capillary pressure $P_c$ between ganglia. Because $P_c$ is proportional to interfacial curvature $\kappa$, and $\kappa$ sets the dissolved concentration at a ganglion's interface by the Kelvin equation, high-$\kappa$ ganglia dissolve more rapidly and feed into low-$\kappa$ ganglia, causing the latter to grow.
In bulk fluids, $\kappa$ is inversely proportional to the bubble radius, which drives smaller bubbles to dissolve while feeding the growth of larger bubbles.
At equilibrium, only one large bubble remains \cite{lifshitz1961orig, wanger1961orig, voorhees1985}. Porous confinement fundamentally alters this picture. For a bubble confined to a single pore, $\kappa$ first \textit{decreases} as the bubble grows until it touches the pore's walls (sub-critical regime), then \textit{increases} as it deforms under the confining geometry (super-critical regime) \cite{xu2017PRL}.
This non-monotonicity admits stable equilibrium of bubble populations that have a multitude of sizes and shapes but equal curvature, impossible in free space \cite{deChalendar2018pnm, mehmani2022JCP}.
In most applications of practical interest, however, ganglia span multiple pores, as evidenced by numerous X-ray imaging studies of rocks undergoing cyclic displacement \cite{iglauer2011gang, goodarzi2024exp, jangda2023pore, boon2024XrayH2}. Here, the relationship between $\kappa$ and ganglion size is far more complex, characterized by oscillatory behavior and multiple discontinuities \cite{wang2021PNAS,mehmani2022AWR}.
Each time a ganglion invades a pore through a narrow throat, $\kappa$ drops abruptly since the ganglion occupies a larger pore volume. As the ganglion grows further, $\kappa$ climbs again.
Growth and shrinkage are therefore punctuated by discrete capillary events such as invasion, snap-off, retraction, fragmentation, coalescence, and dislocation, which alter the ganglion's topology \cite{mehmani2022AWR}. Capturing this physics is key to any predictive pore-scale model of Ostwald ripening.

These rich dynamics have motivated a range of theoretical descriptions at different scales. At Darcy scale, continuum models describe ganglia with macroscopic field variables like saturation \cite{yaxin2020JFM, mehmani2024deplete, salehpour2025micro}, while equilibrium \cite{mehmani2022JCP,bueno2023PNM} and kinetic theories \cite{yu2023GRLkinetics, bueno2024theory, bueno2025theory} describe how the statistics of bubble populations evolve. Pore-scale models play a vital role in formulating and closing such theories.
Among them, direct numerical simulation (DNS) provides detailed mechanistic insight by resolving the fluid-fluid interface in full \cite{singh2022level,singh2024ostwald}. However, the cost of DNS is prohibitive for studying sufficiently large ganglion populations that yield statistically meaningful information. Pore-network models (PNMs) remain the only viable option, which represent the void space by a computational graph comprised of pore bodies connected by throats. Recently for ripening, a series of PNMs with increasing levels of sophistication have been proposed \cite{deChalendar2018pnm, mehmani2022JCP, mehmani2022AWR}, whose validation has relied primarily on controlled microfluidic experiments \cite{xu2017PRL,joewondo2023exp}.

A major gap in the above PNMs for ripening is that they restrict ganglia to a single pore, and the microfluidic experiments used to validate them meet this condition. However as discussed, multi-pore spanning ganglia dominate in practice, and the only PNM to have relaxed this assumption is \cite{mehmani2022AWR}. But this PNM assumes instantaneous capillary equilibration within each ganglion, does not capture fluid flow, and requires all pores to share the same shape to be tractable. The PNM of \cite{mehmani2022AWR} also consists of a delicate and path-dependent root-finding procedure to compute each ganglion's $P_c$, which makes generalization to heterogeneous pore shapes difficult.
A key ingredient in PNMs for ripening is a $\kappa_b$--$S_b$ curve that relates bubble curvature $\kappa_b$ to saturation $S_b$ within each pore. The curve is also vital to most dynamic PNMs of two-phase flow \cite{thompson2002dpnm, joekar2010dpnm, chen2020dpnm}, used to study fluid-fluid displacement. In these PNMs, $\kappa_b$--$S_b$ curves are based on idealized pore geometries (cube, polyhedron) that are amenable to an analytical form.
Moreover, this form is held fixed regardless of the connectivity of the non-wetting phase in the pore, an assumption relaxed only in \cite{mehmani2022AWR}.
Another simplification in displacement-centric PNMs is that the sub-critical branch of the $\kappa_b$--$S_b$ curve is regularized for numerical robustness. This is inconsequential for forced displacement problems, where viscous forces dominate. However for ripening, where the driving force consists of differences in $\kappa_b$, such regularization can prove catastrophic. For example, if $k_b$ is assumed constant on the sub-critical branch \cite{joekar2010dpnm}, spherical bubbles of different sizes would cease to ripen, contrary to experimental evidence \cite{xu2017PRL}. Nevertheless, similar simplifications about capillary pressure within pores have been made in older PNMs used to study ganglion dissolution and pressure exsolution \cite{yortsos1995visual, dillard2000NAPL, dominguez2000gas}.
\looseness=-1

We present an image-based pore-network model (iPNM) that addresses all the above limitations. iPNM operates directly on binarized images of a porous microstructure, extracting its network and computing distinct $\kappa_b$--$S_b$ curves for each pore via the pore-morphology method \cite{hazlett1995pmm, hilpert2001pmm} (modified for ripening \cite{mehmani2024deplete}). These $\kappa_b$--$S_b$ curves apply to arbitrary pore shapes, obviating any geometric simplification.
Moreover, iPNM classifies bubble-occupied pores into three types (singleton, terminal, and bridge), each with a distinct $\kappa_b$--$S_b$ curve that dynamically adapts to the local connectivity of the non-wetting phase.
This divide-and-conquer strategy avoids the analytically intractable task of constructing highly oscillatory capillary pressure versus bubble-size relations for multi-pore ganglia \cite{wang2021PNAS}, their effects instead emerging implicitly from the pore-wise $\kappa_b$--$S_b$ curves.
iPNM couples two-phase flow, solute transport, and Ostwald ripening while handling all capillary events a ganglion may undergo. Unlike the quasi-static PNM of \cite{mehmani2022AWR}, no intricate root finding is needed.
Finally, iPNM establishes a one-to-one mapping between simulated pore saturations on the network's graph and the microstructure image, allowing fast calculations on the graph to be transformed into detailed images of phase distribution similar to DNS.
We validate iPNM against recent microfluidic experiments \cite{salehpour2025micro} of controlled and long-term (15--24 days) measurements of Ostwald ripening in a realistic sandstone microstructure, containing multi-pore H$_2$ ganglia. Very good agreement is observed without any adjustable parameters.

The paper's outline follows: Section~\ref{sec:probdesc} describes the problem and governing physics.  Section~\ref{sec:ipnm} formulates iPNM. Section~\ref{sec:num} verifies iPNM against a previous quasi-static PNM \cite{mehmani2022AWR} in capturing key capillary events. Section~\ref{sec:exp} validates iPNM against microfluidic experiments \cite{salehpour2025micro}. We discuss results in Section~\ref{sec:disc} and summarize findings in Section~\ref{sec:conclusion}.

\section{Problem description}
\label{sec:probdesc}

\begin{figure}[t!]
\centering
\includegraphics[width=\textwidth]{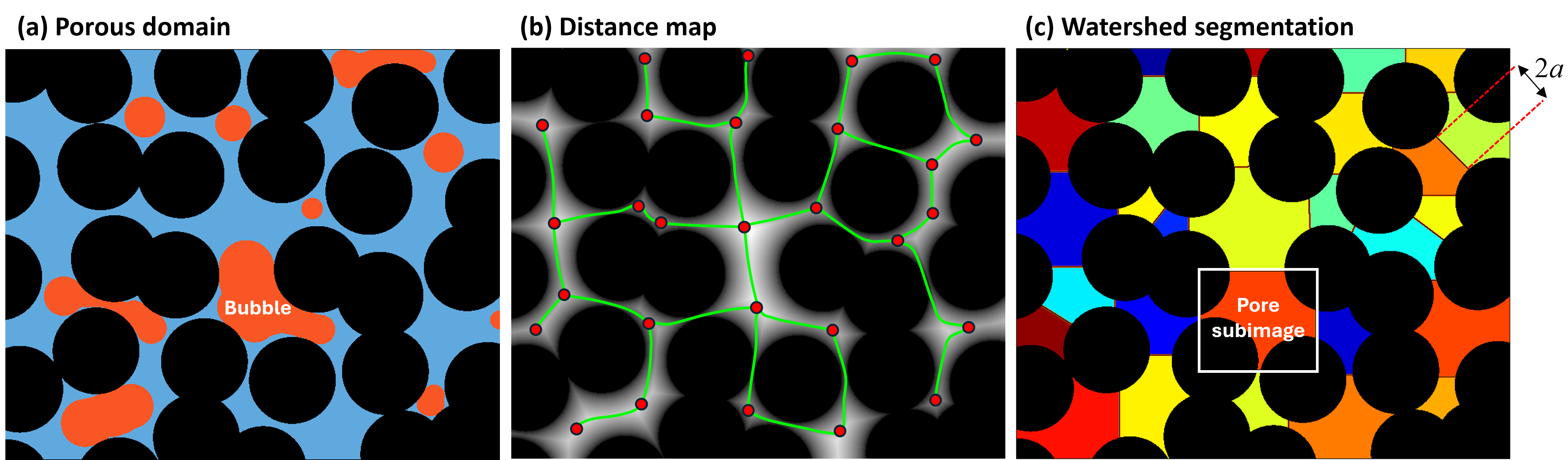}
\caption{(a) Schematic of a porous medium, where solid grains are colored black, trapped ganglia white, and the wetting phase gray. (b) Distance map of the void space with the extracted pore network superposed, represented as a computational graph of nodes (pores) and links (throats). (c) Watershed segmentation partitioning the void space into pores, shown as colored regions. Each node in the network corresponds to a region. The boxed region in (c) illustrates a local pore subimage used to compute the pore-wise $\kappa_b$--$S_b$ curve depicted in Fig.\ref{fig:kbsb}.}
\label{fig:domain}
\end{figure}

We consider a rigid porous medium whose void space is partially occupied by trapped non-wetting ganglia, as shown in Fig.\ref{fig:domain}a. Black represents the solid grains, gray the void space filled with the wetting phase, and white the trapped ganglia. The wetting phase is a liquid in which the gaseous non-wetting phase is partially miscible. Ganglia vary in size and span one or more pores. We assume local thermodynamic equilibrium between each ganglion and the immediately surrounding liquid. The concentration of dissolved non-wetting species at a ganglion's interface is governed by the Kelvin equation.
We also assume bubble composition is uniform and the vapor pressure of the wetting component within each ganglion is constant.
Differences in interfacial curvature $\kappa$ between ganglia engender concentration gradients in the wetting phase, driving molecular diffusion from regions of high $\kappa$ to low $\kappa$. This process is called Ostwald ripening, and it causes high-$\kappa$ ganglia to shrink and low-$\kappa$ ganglia to grow. We also assume that the capillary pressure variations among ganglia are small relative to the ambient pressure magnitude, rendering the non-wetting density approximately spatially uniform. We justify this assumption later for the experiments in Section~\ref{sec:exp}.

Porous confinement fundamentally alters the relationship between $\kappa$ and ganglion size compared to bubbles in a bulk fluid, as discussed in Section~\ref{sec:intro}. As ganglia evolve during ripening, they undergo discrete capillary events that alter their topology. They include invasion (entering a new pore), retraction (withdrawal from a pore), snap-off (pinching of a meniscus in a throat), fragmentation (splitting into disjoint parts), coalescence (merging of disjoint parts), and dislocation (migration into an adjacent pore). Such topological changes redistribute the non-wetting phase and drive local flow within the wetting phase, even in the absence of an external pressure gradient. Capturing this ripening-induced flow is essential, as it governs the timescale and fate of mass redistribution among ganglia.

\section{Image-based pore network model (iPNM)}
\label{sec:ipnm}

\begin{figure}[t!]
\centering
\includegraphics[width=0.5\textwidth]{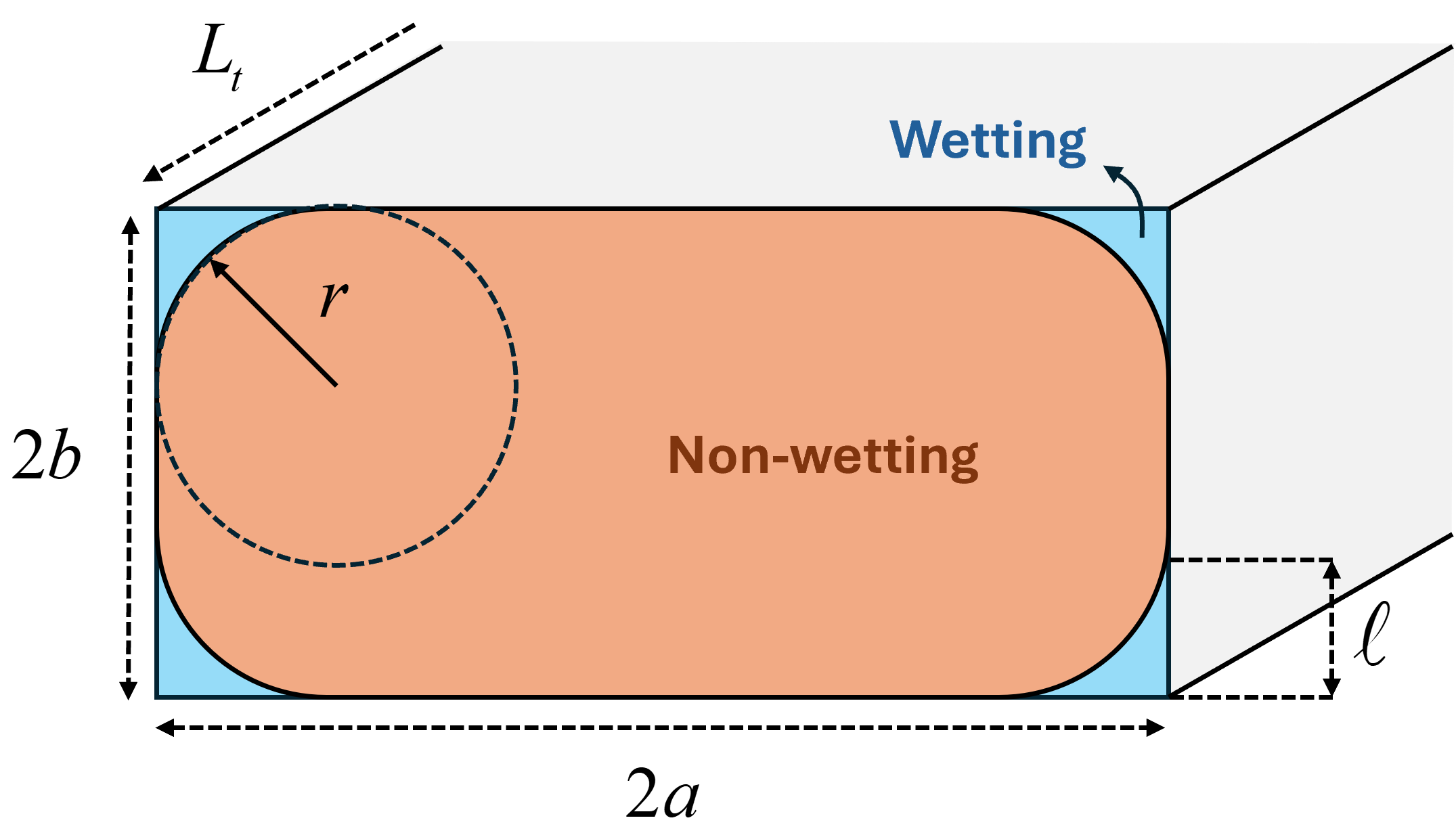}
\caption{Cross-section of a rectangular throat with half-widths $a$ and $b$. Non-wetting phase (orange) occupies the interior, and wetting phase resides in the corners. The radius of curvature of the corner arc menisci (AMs) is $r$ and the distance from the contact line to the corner apex is $\ell$.}
\label{fig:throat_rect}
\end{figure}

\subsection{Network extraction}
\label{sec:ipnm_net}

iPNM operates on a binarized image of a porous microstructure to extract a pore network: a computational graph made up of nodes (pores) and links (throats) representing geometric enlargements and constrictions of the void space, respectively (Fig.\ref{fig:domain}b). 
First, a distance map of the void space is computed, assigning to each void pixel its distance to the nearest solid pixel (Fig.\ref{fig:domain}b). Local maxima of this distance map are then identified and used as markers for watershed segmentation \cite{beucher1979water, gostick2017watershedmarker}, which partitions the void into distinct pores shown by colored regions in Fig.\ref{fig:domain}c. The volume of each pore $V_p$ is the sum of its constituent pixel volumes, and the maximum inscribed radius of each pore $R_p$ is the largest distance-map value in the region.
In iPNM, the local subimage of each pore (e.g., boxed region in Fig.\ref{fig:domain}c) is also retained for computing pore-wise $\kappa_b$--$S_b$ curves, as described in Section~\ref{sec:ipnm_pore}.
Throats linking adjacent pores are approximated as prisms with rectangular cross sections. This is appropriate for the topologically 2D micromodel analyzed in Section~\ref{sec:exp}, and shown to be an accurate representation in prior work \cite{mehmani2017minimum}. The in-plane half-width $a$ of each throat's cross section (Figs.\ref{fig:domain}c and \ref{fig:throat_rect}) is the maximum inscribed radius at the constriction, i.e., pixels along the interface shared between two colored regions (Fig.\ref{fig:domain}c). The out-of-plane half-width $b$ (Fig.\ref{fig:throat_rect}) equals the uniform gap thickness $g$ of the micromodel divided by two. The length of each throat $L_t$ is the sum of the Euclidean distances from the centroid of the constriction to the centroids of the two neighboring pores. We note that iPNM applies equally to networks with 3D topologies as well as other throat geometries.
Throats carry no volume in iPNM, i.e., all void volume is assigned to pores. Throats provide the connectivity between pores and hydraulic conductance for flow.

\subsection{Balance equations}
\label{sec:ipnm_bal}
The image-based pore network model (iPNM) is built upon three balance equations formulated for each pore $i$: a wetting-phase molar balance, a non-wetting (bubble) phase molar balance, and a dissolved-species molar balance. In the following, we present these equations in their most general form, then simplify them by exploiting inherent timescale separations and order-of-magnitude approximations valid in the dilute limit. The result is a set of numerically tractable equations that can be solved sequentially via operator splitting.

Conservation of the wetting phase in pore $i$ reads:
\begin{empheq}[box=\fbox]{equation}
\label{eq:wet}
\rho_w V_i \frac{dS_{w,i}}{dt} = \rho_w \sum_{j=1}^{z_i} q_{w,ij}
\end{empheq}
where $\rho_w$ is the molar density of the wetting phase, $V_i$ is the pore volume, $S_{w,i}$ is the wetting-phase saturation, $z_i$ is the coordination number, and $q_{w,ij}$ is the volumetric flow rate of wetting phase through throat $ij$ (positive into pore $i$). No mass-transfer source appears as the partial pressure of the wetting component in each bubble is assumed constant.

Conservation of the non-wetting component residing in the bubble phase reads:
\begin{equation}
\label{eq:bub}
\rho_b V_i \frac{dS_{b,i}}{dt} = \rho_b \sum_{j=1}^{z_i} q_{b,ij} + \dot{m}_i,
\end{equation}
where $\rho_b$ is the molar density of the bubble phase (treated as constant under the incompressibility assumption; see Section~\ref{sec:probdesc}), $S_{b,i} = 1 - S_{w,i}$ is the bubble saturation, $q_{b,ij}$ is the volumetric flow rate of the bubble phase through throat $ij$, and $\dot{m}_i$ [mol/s] is the interphase mass-transfer rate (positive for exsolution, negative for dissolution).

Lastly, conservation of the dissolved non-wetting species in the wetting phase reads:
\begin{equation}
\label{eq:dis}
\frac{d(\rho_w C_i S_{w,i} V_i)}{dt} = \rho_w \sum_{j=1}^{z_i^w} J_{ij} - \dot{m}_i,
\end{equation}
where $C_i$ is the mole fraction of the dissolved non-wetting species and $J_{ij}$ is the solute flux through throat $ij$ (defined in Section~\ref{sec:ipnm_throat}). The summation runs over the $z_i^w = z_i - z_i^b$ unoccupied throats (filled with wetting phase) connected to pore $i$, where $z_i^b$ denotes the number of bubble-occupied throats ($z_i^b \leq z_i$). We neglect dissolved-species transport through occupied throats because the wetting phase there is confined to thin corner films whose cross-sectional area is much smaller than that of unoccupied throats. Furthermore, hydrodynamic transport of the non-wetting component through occupied throats via the bubble phase redistributes mass within a ganglion much faster than the transport of dissolved species through corner films. This intra-ganglion redistribution is captured by the flow equations (Eqs.\ref{eq:wet}--\ref{eq:bub}).

The right-hand side of Eq.\ref{eq:bub} contains two processes operating on distinct timescales: bubble-phase flow ($q_{b,ij}$; fast) and interphase mass transfer ($\dot{m}_i$; slow). We decouple them via Lie--Trotter operator splitting. Setting $\dot{m}_i = 0$ in Eq.\ref{eq:bub}:
\begin{empheq}[box=\fbox]{equation}
\label{eq:bub_flow}
\rho_b V_i \frac{dS_{b,i}}{dt} = \rho_b \sum_{j=1}^{z_i} q_{b,ij}
\end{empheq}
Eqs.\ref{eq:wet} and \ref{eq:bub_flow}, combined with the constraint $S_w + S_b = 1$, constitute the standard incompressible two-phase \textit{flow problem}, which we solve for $P_{w,i}$ and $S_{b,i}$ while holding $C_i$ frozen.

Setting $q_{b,ij} = 0$ in Eq.\ref{eq:bub} yields:
\begin{equation}
\label{eq:bub_trans}
\rho_b V_i \frac{dS_{b,i}}{dt} = \dot{m}_i.
\end{equation}
By adding Eqs.\ref{eq:bub_trans} and \ref{eq:dis}, we can eliminate $\dot{m}_i$ and obtain the following total non-wetting component balance:
\begin{equation}
\label{eq:total}
\rho_b V_i \frac{dS_{b,i}}{dt} + \frac{d(\rho_w C_i S_{w,i} V_i)}{dt} = \rho_w \sum_{j=1}^{z_i^w} J_{ij}.
\end{equation}
For a vacant pore ($S_{b,i} = 0$), the first term on the left-hand side of Eq.\ref{eq:total} vanishes, yielding the standard advection--diffusion equation for $C_i$:
\begin{empheq}[box=\fbox]{equation}
\label{eq:vac}
\rho_w V_i \frac{dC_i}{dt} = \rho_w \sum_{j=1}^{z_i^w} J_{ij}
\end{empheq}

For an occupied pore ($S_{b,i} > 0$), we simplify Eq.\ref{eq:total} by first expanding the second term on the left-hand side:
\begin{equation}
\label{eq:expand}
\frac{d(\rho_w C_i S_{w,i} V_i)}{dt} = \rho_w V_i \left( S_{w,i} \frac{dC_i}{dt} + C_i \frac{dS_{w,i}}{dt} \right).
\end{equation}
Because the dissolved molar concentration $\rho_w C_i$ is much smaller than the molar density in the bubble phase $\rho_b$, we neglect the second term on the right-hand side of Eq.\ref{eq:expand}. Moreover, the concentration $C_i$ in an occupied pore is determined by local thermodynamic equilibrium: $C_i$ depends on capillary pressure $P_{c,i}$, which in turn depends on the bubble saturation $S_{b,i}$ (these functional relationships are presented in Section~\ref{sec:ipnm_pore}). Applying the chain rule gives:
\begin{equation}
\label{eq:chain}
\frac{dC_i}{dt} = \frac{dC}{dP_c} \frac{dP_c}{dS_b} \frac{dS_{b,i}}{dt}.
\end{equation}
Substituting Eq.\ref{eq:chain} into Eq.\ref{eq:total}, the prefactor $\rho_w S_w (dC/dP_c)(dP_c/dS_b)$ multiplying the first term on the right-hand side of Eq.\ref{eq:expand} becomes negligible compared to $\rho_b$ for our H$_2$--H$_2$O system. This follows from an order-of-magnitude approximation detailed in Remark 1 of Section~\ref{sec:ipnm_pore}. Hence, we can drop the foregoing term, reducing Eq.\ref{eq:total} to:
\begin{empheq}[box=\fbox]{equation}
\label{eq:occ}
\rho_b V_i \frac{dS_{b,i}}{dt} = \rho_w \sum_{j=1}^{z_i^w} J_{ij}
\end{empheq}
Eq.\ref{eq:occ} states that the rate of bubble growth is driven entirely by the net solute flux $J_{ij}$ through unoccupied throats.

In summary, the balance equations solved in iPNM include Eq.\ref{eq:wet} for wetting-phase conservation, Eq.\ref{eq:bub} for non-wetting phase conservation, and Eq.\ref{eq:dis} for dissolved-species conservation. The \textit{flow problem} consists of Eqs.\ref{eq:wet} and \ref{eq:bub_flow} and is solved first for $P_{w,i}$ and $S_{b,i}$ while holding $C_i$ frozen.
The \textit{transport problem} consists of Eqs.\ref{eq:occ} and \ref{eq:vac} and is solved second for $C_i$ and $S_{b,i}$ while holding $q_{w,ij}$ frozen (computed in the flow problem).
A capillary stability check---encompassing invasion, snap-off, and retraction events---is performed after each problem as described in Section~\ref{sec:ipnm_cap}.

\subsection{Throat constitutive equations}
\label{sec:ipnm_throat}

We next present constitutive equations for throats with rectangular cross sections. The geometric parameters $a$, $b$, and $L_t$ were defined in Section~\ref{sec:ipnm_net}. Recall that for a micromodel with uniform gap thickness $g$, we have $b = g/2$.

\medskip
\noindent\textbf{Flux laws.}
For Newtonian fluids at low Reynolds number, the flow rate of phase $\alpha \in \{w, b\}$ through throat $ij$ is:
\begin{equation}
\label{eq:flux_q}
q_{\alpha,ij} = \frac{K_{\alpha,ij}}{\mu_\alpha}(P_{\alpha,j} - P_{\alpha,i}),
\end{equation}
where $K_{\alpha,ij}$ is the hydraulic conductivity of phase $\alpha$ and $\mu_\alpha$ is its dynamic viscosity. The net solute flux through an unoccupied throat $ij$, which appears in Eq.\ref{eq:dis}, comprises advective transport and Fickian diffusion:
\begin{equation}
\label{eq:flux_J}
J_{ij} = q_{w,ij}\,C_{ij}^{\mathrm{up}} + D_m\frac{A_{ij}}{L_{ij}}(C_j - C_i),
\end{equation}
where $C_{ij}^{\mathrm{up}}$ is the upwind pore's mole fraction, $D_m$ is the molecular diffusion coefficient, and $A_{ij}$ and $L_{ij}$ are the cross-sectional area and length of the throat. Below, we replace the $ij$ subscript with $t$ for throat properties (e.g., $A_t$, $L_t$).

\medskip
\noindent\textbf{Single-phase conductivity.}
Throats that are not invaded by the non-wetting phase are fully occupied by the wetting phase.
For an unoccupied throat, $K_b = 0$ and $K_w = K_t$, where $K_t$ is the single-phase conductivity. For a rectangular cross-section, the analytical solution to the Stokes equations yields \cite{white2006book, patzek2001CondOneS}:
\begin{equation}
\label{eq:Kt}
K_t = \tilde{g}(\varepsilon)\,\frac{A_t^2}{L_t},
\end{equation}
where $A_t = 4\,ab$ is the throat's cross-sectional area, $L_t$ its length, and:
\begin{equation}
\label{eq:gtilde}
\tilde{g}(\varepsilon) = \left(1 + \frac{1}{\varepsilon}\right)^2\left[\frac{1}{3} - \frac{64}{\varepsilon\pi^5}\sum_{n=0}^{\infty}\frac{\tanh[(2n+1)\pi\varepsilon/2]}{(2n+1)^5}\right]G_t
\end{equation}
the dimensionless conductivity. The aspect ratio and shape factor of the throat are computed via $\varepsilon = a/b$ and $G_t = A_t/P_t^2$, respectively, where $P_t \!=\! 4(a+b)$ is the throat's perimeter. Eqs.\ref{eq:Kt}--\ref{eq:gtilde} are exact, but approximate closed-form expressions are also available \cite{muzychka1998CondOne}. For a square cross-section, Eq.\ref{eq:gtilde} reduces to $\tilde{g} = 0.5623\,G_t$.
Mehmani and Tchelepi \cite{mehmani2017minimum} found that Eq.\ref{eq:Kt} yields network permeability predictions within 10\% of direct numerical simulation.

\medskip
\noindent\textbf{Two-phase wetting conductivity.}
Throats invaded by the non-wetting phase contain both a central bubble region and wetting films occupying the four corners. The wetting-phase conductivity $K_w$ (Eq.\ref{eq:flux_q}) is computed following \cite{patzek2001CondOneK}:
\begin{equation}
\label{eq:Kw}
K_w = \frac{8\,\tilde{g}_w\,\ell^4}{L_t},
\end{equation}
where $\ell \!=\! r\cos(\theta+\beta)/\sin\beta$ is the distance from the corner arc meniscus (AM) to the throat corner (Fig.\ref{fig:throat_rect}), $\beta = \pi/4$ is the corner half-angle, and $r = \sigma/P_c$ is the AM's radius of curvature, with $\sigma$ denoting the interfacial tension. The:
\begin{equation}
\label{eq:gw}
\tilde{g}_w = \exp\left[\frac{a_1\tilde{G}_w^2 + a_2\tilde{G}_w + a_3 + 0.02\sin(\beta-\pi/6)}{(1/4\pi - \tilde{G}_w)^{e_1}\cos^{e_2}(\beta-\pi/6)} + 2\ln\tilde{A}_w\right],
\end{equation}
is the dimensionless conductivity of one half-corner film, with dimensionless cross-sectional area:
\begin{equation}
\label{eq:Aw_dimless}
\tilde{A}_w = \left(\frac{\sin\beta}{\cos(\theta+\beta)}\right)^2\left[\frac{\cos\theta\cos(\theta+\beta)}{\sin\beta} + \theta + \beta - \frac{\pi}{2}\right],
\end{equation}
and corresponding shape factor:
\begin{equation}
\label{eq:Gw}
\tilde{G}_w = \frac{\tilde{A}_w}{4\left[1 - \left(\theta + \beta - \frac{\pi}{2}\right){\sin\beta}\,/\,{\cos(\theta+\beta)}\right]^2}.
\end{equation}
The $e_1$, $e_2$, $a_1$, $a_2$, and $a_3$ are fitting coefficients. In \cite{patzek2001CondOneK}, these are provided for no-slip and perfect-slip conditions at the bubble-water interface. We use the perfect-slip coefficients $(e_1, e_2, a_1, a_2, a_3) = (1, 0, -18.2066, 5.88287, -0.351809)$, appropriate for the gaseous H$_2$ bubbles immersed in water as considered in the experiments of Section \ref{sec:exp}.

\medskip
\noindent\textbf{Two-phase non-wetting conductivity.}
The non-wetting phase conductivity $K_b$ (Eq.\ref{eq:flux_q}) for a rectangular throat is not available in the literature. We derive one here using the hydraulic diameter concept \cite{white2006book}, ensuring that $K_b \to K_t$ as $S_b \to 1$. The derivation is given in \ref{app:cond}. Here we present the final result:
\begin{equation}
\label{eq:Kb}
K_b = \frac{D_{h,b}^2\,A_b}{2\,{Po}\,L_t},
\end{equation}
where $A_b \!=\! A_t - A_w$ is the cross-sectional area occupied by the non-wetting phase, with $A_w = 4\,\ell^2\tilde{A}_w$ the total wetting-phase area in the four corners ($\ell$ and $\tilde{A}_w$ were defined previously).
The hydraulic diameter of the bubble region is $D_{h,b} = 4\,A_b/C_b$, where the wetted perimeter, invoking again perfect slip at the bubble-water interface, is:
\begin{equation}
\label{eq:Cb}
C_b = 4(a - \ell) + 4(b - \ell).
\end{equation}
The Poiseuille number $Po$ is computed from the single-phase conductivity:
\begin{equation}
\label{eq:Po}
{Po} = \frac{D_h^2\,A_t}{2\,K_t\,L_t},
\end{equation}
and assumed to be independent of saturation. Here, $D_h = 4\,A_t/P_t$ is the hydraulic diameter of the throat itself.

\medskip
\noindent\textbf{Capillary entry pressure.}
We apply the Mayer-Stowe-Princen (MS-P) theory \cite{mayer1965MSP,princen1969MSP} to derive the capillary entry pressure of throats with rectangular cross-sections. MS-P equates the curvature of the AMs in the corners of the throat to the curvature of the main terminal meniscus (MTM)---the advancing part of the interface. The entry pressure is:
\begin{equation}
\label{eq:Pce}
P_{ce} = \frac{\sigma}{r},
\end{equation}
where $r$ is the AM radius of curvature. The latter is obtained by solving:
\begin{equation}
\label{eq:MSP}
\frac{1}{r} = \frac{P_e(r)}{A_e(r)},
\end{equation}
where $P_e$ is the effective perimeter and $A_e$ the effective cross-sectional area of the non-wetting phase, both functions of $r$. Explicit expressions for $P_e(r)$ and $A_e(r)$ are derived in \ref{app:entry} for a rectangular throat with side lengths $2a$ and $2b$. Eq.\ref{eq:MSP} is solved numerically using the bisection method. For the case of a throat with square cross-section ($a \!=\! b$), the derived expressions reduce to the analytical solution by \cite{ma1996PceSquare}. The proof is given in \ref{app:entry}.

\medskip
\noindent\textbf{Snap-off pressure.}
Snap-off occurs when the AMs in adjacent corners of a throat coalesce, at which point any infinitesimal decrease in capillary pressure becomes unsustainable, causing the wetting phase to fill the throat. For a rectangular throat with half-widths $a$ and $b$, snap-off occurs when the distance from each AM's contact line to the corner of the throat equals $\min(a,b)$ (see \ref{app:entry}), giving the snap-off pressure:
\begin{equation}
\label{eq:pcs}
P_{cs} = \frac{\sigma \sin(\pi/4 - \theta)}{\min(a,b) \sin(\pi/4)}.
\end{equation}
For a square cross-section, this reduces to $P_{cs} = \sigma\,(\cos\theta - \sin\theta)/a$, as proposed by Ma et al. \cite{ma1996PceSquare}. For zero contact angle in a rectangular throat, Eq.\ref{eq:pcs} reduces to $P_{cs} = \sigma/\min(a,b)$, which was proposed by Lenormand et al. \cite{lenormand1983Pcs}.

\subsection{Pore constitutive equations}
\label{sec:ipnm_pore}
We next present constitutive equations for pores. The pore volume $V_p$ and maximum inscribed radius $R_p$ were defined in Section~\ref{sec:ipnm_net}. Beyond these scalar quantities, iPNM requires the local subimage of each pore to compute its $\kappa_b$--$S_b$ curve via the pore morphology method (PMM). Each subimage corresponds to a colored region in Fig.\ref{fig:domain}c.

\begin{figure}[t!]
\centering
 \includegraphics[width=\textwidth, trim=15 15 20 20, clip]{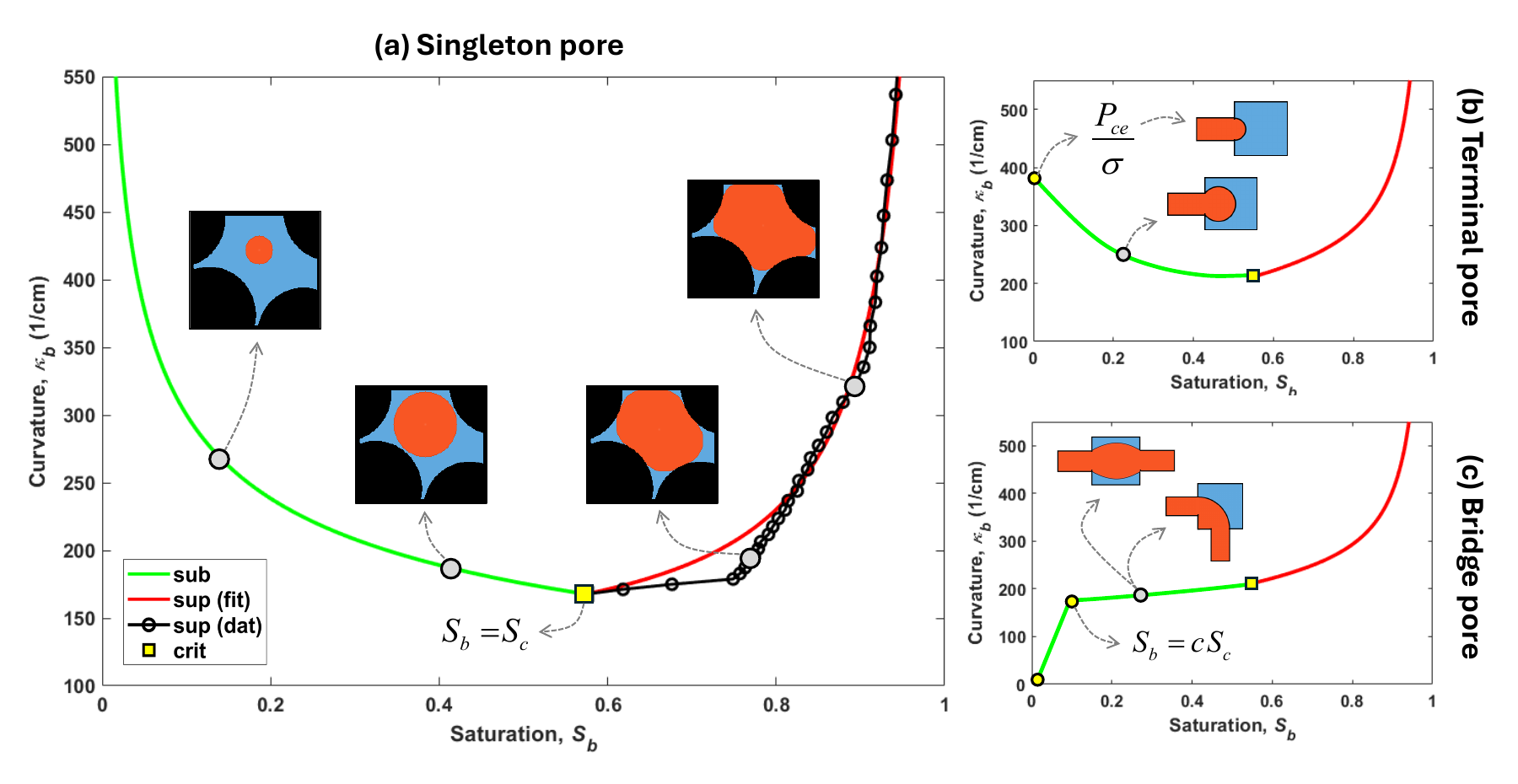}
\caption{Bubble curvature--saturation ($\kappa_b$--$S_b$) relationships for different pore types. (a) Singleton pore: the sub-critical branch (green) corresponds to a spherical bubble that does not touch the pore walls; the super-critical branch (red) is fitted to PMM data (black circles); the yellow square marks the critical point ($S_c$, $\kappa_{b,\min}$). Insets show the bubble configuration at different saturations obtained from PMM. (b) Terminal pore: the sub-critical branch (green) intersects the $\kappa_b$-axis at $P_{ce}/\sigma$, corresponding to the entry pressure of the connected invaded throat. Insets show the bubble at the moment of invasion and shortly after. (c) Bridge pore: the interface emerges from one throat and terminates at another, allowing curvatures below $\kappa_{b,\min}$. The linear sub-critical branch ensures $\kappa_b \to 0$ as $S_b \to 0$ (flat interface). Insets show possible interface configurations.}
\label{fig:kbsb}
\end{figure}

\medskip
\noindent\textbf{Capillary-pressure curve.}
The capillary pressure $P_c$ of a bubble-occupied pore depends on the mean interfacial curvature $\kappa_b$ of the non-wetting phase within that pore through the Young-Laplace equation:
\begin{equation}
\label{eq:Pc_kb}
P_c = \sigma \kappa_b,
\end{equation}
For micromodels with uniform gap thickness $g$ and zero contact angle, as those in Section~\ref{sec:exp}, Eq.\ref{eq:Pc_kb} reduces to:
\begin{equation}
\label{eq:kb_2D}
\kappa_b = \kappa_b^{ip} + \kappa_{gap}, \quad \kappa_{gap} = 2/g,
\end{equation}
where $\kappa_b^{ip}$ is the in-plane curvature and $\kappa_{gap}$ is the constant out-of-plane curvature.
Mean curvature $\kappa_b$ (or $\kappa_b^{ip}$ in 2D) is a direct function of the pore's saturation $S_b = V_b/V_p$, where $V_b$ is the non-wetting phase volume in the pore.
The specific form of this $\kappa_b$--$S_b$ relation depends on the \textit{pore type}, which we classify into three categories on the basis of how many invaded throats they connect to ($z^b$): singleton ($z^b = 0$), terminal ($z^b = 1$), and bridge ($z^b > 1$) \cite{mehmani2022AWR}. 
These are depicted in Fig.\ref{fig:kbsb} along with their corresponding  $\kappa_b$--$S_b$ curves, which we discuss next for each pore type.

\medskip
\noindent\textit{Singleton.}
For a pore containing a single-pore spanning bubble (Fig.\ref{fig:kbsb}a):
\begin{equation}
\label{eq:kbsb_sng}
\kappa_b = \begin{cases} \kappa_{b,\min}\left(\dfrac{S_c}{S_b}\right)^{1/D}, & S_b < S_c \\[12pt] \kappa_{b,\min}\left(1 + \lambda_1\dfrac{S_b - S_c}{1 - S_c} + \lambda_2\dfrac{S_b - S_c}{1 - S_b}\right), & S_b \geq S_c \end{cases}
\end{equation}
where $D$ is the network's dimensionality ($=2$ for micromodels). The \textit{super-critical} branch ($S_b \geq S_c$) corresponds to bubbles that are deformed by and conform to the pore's walls. The \textit{sub-critical} branch ($S_b < S_c$) corresponds to bubbles that do not touch the walls and remain spherical (or circular in 2D). The minimum curvature $\kappa_{b,\min} = 2/R_p$ (or $1/R_p$ in 2D) occurs at the critical saturation $S_c$, where $R_p$ is the maximum inscribed radius. The \textit{critical} saturation is $S_c = V_c/V_p$, where $V_c = 4\pi R_p^3/3$ in 3D (or $V_c = \pi R_p^2 g$ in 2D).
The parameters $\lambda_1$ and $\lambda_2$ depend on the pore shape.

\medskip
\noindent\textit{Terminal.}
For a pore that is part of a multi-pore ganglion, with one invaded throat of entry pressure $P_{ce}$ (Fig.\ref{fig:kbsb}b):
\begin{equation}
\label{eq:kbsb_ter}
\kappa_b = \begin{cases} \left(\dfrac{P_{ce}}{\sigma} - \kappa_{b,\min}\right)\left(\dfrac{S_c - S_b}{S_c}\right)^n + \kappa_{b,\min}, & S_b < S_c \\[12pt] \kappa_{b,\min}\left(1 + \lambda_1\dfrac{S_b - S_c}{1 - S_c} + \lambda_2\dfrac{S_b - S_c}{1 - S_b}\right), & S_b \geq S_c \end{cases}
\end{equation}
The super-critical branch is identical to Eq.\ref{eq:kbsb_sng}, but the sub-critical branch differs. This is because at the instant when the non-wetting phase invades a terminal pore (i.e., the terminal pore is ``born''), capillary pressure equals $P_{ce}$ (Fig.\ref{fig:kbsb}b). Hence, as $S_b\rightarrow 0$, we must have $\kappa_b \rightarrow P_{ce}/\sigma$. The parameter $n$ is empirical, which we set to $n = 2$ for smoothness.

\medskip
\noindent\textit{Bridge.}
For a pore that is part of multi-pore ganglion, and connected to multiple invaded throats (Fig.\ref{fig:kbsb}c):
\begin{equation}
\label{eq:kbsb_brg}
\kappa_b = \begin{cases} \kappa_{b,c}\dfrac{S_b}{c\,S_c}, & S_b < c\,S_c \\[12pt] \kappa_{b,\min}\left(1 + \lambda_1\dfrac{S_b - S_c}{1 - S_c} + \lambda_2\dfrac{S_b - S_c}{1 - S_b}\right), & S_b \geq c\,S_c \end{cases}
\end{equation}
The second case ($S_b\geq c\,S_c$) is identical to the super-critical branch of Eq.\ref{eq:kbsb_sng}. The first case is a line crossing the origin $(S_b,\kappa_b) = (0,0)$.
The rationale is that in a bridge pore, the interface emanates from one throat and terminates at another, forming a curved surface \textit{through} the pore whose curvature can fall well below $\kappa_{b,\min}$ (Fig.\ref{fig:kbsb}c).
Complete flattening of the interface, however, never occurs because one or more of the connected throats would snap off first, transforming the bridge pore into a terminal or singleton.
Eq.\ref{eq:kbsb_brg} therefore extrapolates the super-critical formula from $S_c$ to $c\,S_c$, then connects it to the origin to ensure $\kappa_b\rightarrow 0$ as $S_b \rightarrow 0$.
The parameter $c$ is evaluated as follows:
\begin{equation}
\label{eq:c_safeguard}
c = \begin{cases} 1, & c^* > 0.1 \\ 0.1, & c^* \leq 0.1 \end{cases}
\end{equation}
where $\kappa_{b,c} \!=\! \kappa(c\,S_c)$ and $c^*$ solves $\kappa_b(c^* S_c) \!=\! 0$, with $\kappa_b(\cdot)$ corresponding to the super-critical formula. For most pore shapes, $c^*\leq 0.1$ and $c = 0.1$ suffices. Rarely $c^* > 0.1$, implying $\kappa_b$ is zero at $S_b>0.1\,S_c$, so we anchor the sub-critical branch at $S_c$ by setting $c = 1$. While Eq.\ref{eq:kbsb_brg} for $S_b \!<\! S_c$ is empirical, its exact form plays a minor role in simulations.
\vspace{0.2cm}

The super-critical branches of Eqs.\ref{eq:kbsb_sng}--\ref{eq:kbsb_brg} are identical and depend on pore shape through the parameters $\lambda_1$ and $\lambda_2$. A key novelty of iPNM is to use the pore morphology method (PMM), adapted for Ostwald ripening by \cite{mehmani2024deplete,
salehpour2025micro}, to compute $\lambda_1$ and $\lambda_2$. Doing so avoids making simplifying assumptions about pore shape, allowing accurate upscaling of its impact on capillary pressure. In applying PMM, each pore's subimage $I_p$ is subjected to morphological opening $I_p\circ B_r$ by a spherical (or circular in 2D) structuring element $B_r$
of radius $r$. By varying $r$ from zero to $R_p$, we get a sequence of images that show the ganglion's configuration in the pore (insets in Fig.\ref{fig:kbsb}a). Associated with each image is a curvature $\kappa_b = 2/r$ (or $\kappa_b^{ip} = 1/r$ in 2D) and a saturation $S_b$, which constitutes the super-critical branch of the $\kappa_b$--$S_b$ curve (black circles in Fig.\ref{fig:kbsb}a). The parameters $\lambda_1$ and $\lambda_2$ are fitted to this data (red line in Fig.\ref{fig:kbsb}a). Notice $S_c$ and $\kappa_{b,\min}$ correspond to $r = R_p$. The super-critical branch (red line in Fig.\ref{fig:kbsb}a) has a vertical asymptote: as $S_b \rightarrow 1$, $\kappa_b\rightarrow\infty$. This limit is never approached in practice because the ganglion invades a vacant throat in its periphery if $\kappa_b$ is sufficiently high.
As ganglia evolve during simulation, a pore's type and thus $\kappa_b$--$S_b$ curve changes dynamically.

\medskip
\noindent\textbf{Equilibrium concentration.}
The dissolved concentration of the non-wetting species in the wetting phase of a pore is governed by local thermodynamic equilibrium with the bubble phase. For a dilute solution, Henry's law reads:
\begin{equation}
\label{eq:henry}
C = \frac{P_b - P_v}{H} = \frac{P_w + P_c(S_b) - P_v}{H},
\end{equation}
where $C$ is the equilibrium mole fraction of the dissolved solute, $P_b$ is the bubble-phase pressure, $P_w$ is the wetting-phase pressure, and $P_c$ is the capillary pressure in the pore. Moreover, $P_v$ is the spatially constant vapor pressure of the wetting phase in the bubble, and $H$ is Henry's constant. $P_c$ depends on bubble saturation $S_b$ through Eqs.\ref{eq:Pc_kb}--\ref{eq:kbsb_brg}.
\\

\medskip
\noindent\textit{Remark 1.}
We are now in a position to justify neglecting the term $\rho_w S_w (dC/dP_c)(dP_c/dS_b)$ compared to $\rho_b$ in Eq.\ref{eq:occ}. For the H$_2$--H$_2$O system of interest here, $dC/dP_c = 1/H$ from Eq.\ref{eq:henry}. The slope $dP_c/dS_b$ is on the order of the capillary pressure $P_c \sim \sigma/R_p$. For typical micromodel pore sizes ($R_p \sim 50$--$100$ $\mu$m) and $\sigma \approx 0.07$ N/m, this gives $dP_c/dS_b \sim 10^3$--$10^4$ Pa. With $H \approx 7.8 \times 10^9$ Pa for H$_2$ at 25$^\circ$C and $\rho_w \approx 55.5 \times 10^3$ mol/m$^3$, the term evaluates to $\rho_w S_w (dC/dP_c)(dP_c/dS_b) \sim 10^{-2}$--$10^{-1}$ mol/m$^3$, which is orders of magnitude smaller than $\rho_b \approx 40$ mol/m$^3$ at 1 atm.

\subsection{Capillary stability}
\label{sec:ipnm_cap}

As the balance equations of Section~\ref{sec:ipnm_bal} evolve $S_{b,i}$ during a simulation, the occupancy of pores and throats by the non-wetting phase may become unstable, inducing capillary events such as snap-off, invasion, retraction, and ganglion coalescence and fragmentation. At each time step, we check whether the current configuration is stable and, if not, trigger the appropriate capillary events and update the occupancy until a stable configuration is found.
Let $\chi_i \in \{0,1\}$ and $\chi_{ij} \in \{0,1\}$ be indicator variables for the occupancy of pore $i$ and throat $ij$ by the non-wetting phase, respectively. To evaluate the capillary stability of a throat, we require the throat's capillary pressure, defined here as:
\begin{equation}
\label{eq:Pct}
P_{c,ij} = \frac{P_{c,i} + P_{c,j}}{2},
\end{equation}
where $P_{c,i}$ and $P_{c,j}$ are computed from the $\kappa_b$--$S_b$ curves (Eqs.\ref{eq:kbsb_sng}--\ref{eq:kbsb_brg}).
Both pores $i$ and $j$ straddling the occupied throat are themselves occupied, so Eq.\ref{eq:Pct} is always well-defined.
We check for three types of capillary events:

\medskip
\noindent\textbf{Snap-off.}
An occupied throat snaps off when its capillary pressure drops below the snap-off pressure (Eq.\ref{eq:pcs}):
\begin{equation}
\label{eq:event_snap}
P_{c,ij} < P_{cs,ij} \;\;\text{and}\;\; \chi_{ij} = 1 \quad\Longrightarrow\quad \chi_{ij} = 0.
\end{equation}
Snap-off fills the throat with wetting phase but leaves the occupancy and saturation of neighboring pores intact.

\medskip
\noindent\textbf{Retraction.}
The non-wetting phase in an occupied pore retracts when its saturation reaches zero:
\begin{equation}
\label{eq:event_retract}
S_{b,i} < \epsilon \;\;\text{and}\;\; \chi_i = 1 \quad\Longrightarrow\quad \chi_i = 0,\;\; S_{b,i} = 0,\;\; \chi_{ij} = 0 \;\;\forall\,j \in \mathcal{N}_i^b,
\end{equation}
where $\epsilon \approx 10^{-16}$ is used to avoid excessively small adaptive time steps (discussed in Section~\ref{sec:ipnm_alg}). $\mathcal{N}_i^b$ is the set of occupied throats connected to pore $i$. Upon retraction, all occupied throats connected to pore $i$ also become vacant.

\medskip
\noindent\textbf{Invasion.}
An unoccupied throat is invaded when the capillary pressure of a neighboring pore at super-critical state ($S_{b} \geq S_{c}$) exceeds the entry pressure of the throat (Eq.\ref{eq:Pce}):
\begin{equation}
\label{eq:event_invade}
\max(P_{c,i},\, P_{c,j}) > P_{ce,ij} \;\;\text{and}\;\; \chi_{ij} = 0 \quad\Longrightarrow\quad \chi_{ij} = 1.
\end{equation}
Only super-critical pores contribute to the maximum in Eq.\ref{eq:event_invade}. A sub-critical bubble does not contact the walls of its confining pore and, therefore, cannot invade a throat. If the invaded throat connects to a previously vacant pore $j$, that pore also becomes occupied ($\chi_j = 1$) and is initialized to $S_{b,j} = 10^{-6}$. Similar to retraction, the initialization avoids excessively small time steps. To conserve mass, the corresponding non-wetting-phase volume ($10^{-6}\,V_j$) is subtracted from the source pore $i$ that drove the invasion. This subtraction is bounded so that $S_{w,i}$ remains in $[0,1]$.
\vspace{0.2cm}

Capillary events can trigger one another by altering the occupied coordination number $z_i^b$ of pores. For example, snap-off reduces $z_i^b$, which can transform a bridge pore ($z_i^b > 1$) into a terminal ($z_i^b = 1$) or a terminal into a singleton ($z_i^b = 0$). Since each pore type has a distinct $\kappa_b$--$S_b$ curve (Section~\ref{sec:ipnm_pore}), the pore's capillary pressure changes, which may cause other throats to undergo snap-off or retraction. Conversely, invasion increases $z_i^b$, which can transform a singleton into a terminal or a terminal into a bridge, thereby altering the local capillary pressure and triggering other events. Finding a stable configuration is therefore iterative. Here, we sweep through all pores and throats, check for all three events, then update $\chi_i$, $\chi_{ij}$, and $S_{b,i}$ accordingly. Once a pore or throat has been processed in a given sweep, it is not re-examined in subsequent sweeps within the same call. This prevents stagnation due to two or more events undoing one another (e.g., a newly invaded pore immediately flagged for retraction). Iterations terminate when no occupancy changes occur between consecutive sweeps. Algorithm~\ref{alg:capequil} summarizes the steps described above.

\begin{algorithm}[t]
\caption{Capillary stability check}\label{alg:capequil}
\begin{algorithmic}[0]
\Statex \textbf{Input:} $S_{b,i}$, $\chi_i$, $\chi_{ij}$
\Statex \textbf{Output:} $S_{b,i}$, $\chi_i$, $\chi_{ij}$, $P_{c,i}$, $P_{c,ij}$
\vspace{2pt}
\State Mark all pores and throats as unprocessed
\Repeat
  \State Compute pore capillary pressures $P_{c,i}$ from Eqs.\ref{eq:kbsb_sng}--\ref{eq:kbsb_brg}
  \State Compute throat capillary pressures $P_{c,ij}$ from Eq.\ref{eq:Pct}
  \Statex \algai \textbf{Snap-off:} For each unprocessed throat with $\chi_{ij} = 1$,
  \State \algai if $P_{c,ij} < P_{cs,ij}$, set $\chi_{ij} = 0$ and mark as processed
  \Statex \algai \textbf{Retraction:} For each unprocessed pore with $\chi_i = 1$,
  \State \algai if $S_{b,i} < \varepsilon$, set $\chi_i = 0$, $S_{b,i} = 0$, $\chi_{ij} = 0\;\forall\,j \in \mathcal{N}_i^b$, and mark pore and connected throats as processed
  \Statex \algai \textbf{Invasion:} For each unprocessed throat with $\chi_{ij} = 0$,
  \State \algai set $P_{c,i}^* = P_{c,i}$ if $S_{b,i} \geq S_{c,i}$, else $P_{c,i}^* = 0$ (same for pore $j$)
  \State \algai \begin{minipage}[t]{0.9\textwidth}if $\max(P_{c,i}^*,\, P_{c,j}^*) > P_{ce,ij}$, set $\chi_{ij} = 1$. If a neighboring pore (say $j$) is vacant, set it to occupied with $S_{b,j} = 10^{-6}$ and subtract $10^{-6}\,V_j$ from the source pore. Mark the invaded throat and pore as processed\end{minipage} \vspace{2pt}
\Until{$\chi_i$ and $\chi_{ij}$ are unchanged}
\end{algorithmic}
\end{algorithm}

\subsection{Discretization and algorithm}
\label{sec:ipnm_alg}
In this section, we describe how the balance equations of Section~\ref{sec:ipnm_bal} are discretized in time and solved algorithmically. We employ backward Euler for time integration and Lie--Trotter operator splitting (introduced in Section~\ref{sec:ipnm_bal}) to decouple the \textit{flow problem} (Eqs.\ref{eq:wet} and \ref{eq:bub_flow}), solved for $P_{w,i}$ and $S_{b,i}$ while holding $C_i$ frozen, and the \textit{transport problem} (Eqs.\ref{eq:vac} and \ref{eq:occ}), solved for $C_i$ and $S_{b,i}$ while holding $q_{w,ij}$ frozen. The two sub-problems are solved sequentially, not iteratively, and accuracy is controlled by an adaptive time step $\Delta t$. Within each outer time step, the flow and transport solvers may independently progress in sub-steps with smaller increments $\delta t \leq \Delta t$. A capillary stability check (Algorithm~\ref{alg:capequil}) is performed after each sub-step of both solvers. Algorithm~\ref{alg:ipnm} summarizes the overall procedure.

\begin{algorithm}[t]
\caption{Image-based pore network model (iPNM)}\label{alg:ipnm}
\begin{algorithmic}[0]
\Statex \textbf{Input:} $P_{w,i}$, $S_{b,i}$, $C_i$, $\chi_i$, $\chi_{ij}$, $T$ (simulation time)
\Statex \textbf{Output:} $P_{w,i}$, $S_{b,i}$, $C_i$, $\chi_i$, $\chi_{ij}$
\vspace{2pt}
\State $\mathrm{time} = 0$
\While{$\mathrm{time} < T$}
  \State $[P_{w,i},\, S_{b,i},\, \chi_i,\, \chi_{ij},\, q_{w,ij}]$ = Flow($P_{w,i}$, $S_{b,i}$, $\chi_i$, $\chi_{ij}$, $\Delta t$) \hfill [Algorithm~\ref{alg:flow}]
  \State $[C_i,\, S_{b,i},\, \chi_i,\, \chi_{ij}]$ = Transport($C_i$, $S_{b,i}$, $q_{w,ij}$, $\chi_i$, $\chi_{ij}$, $\Delta t$) \hfill [Algorithm~\ref{alg:transport}]
  \State Adapt $\Delta t$
  \State $\mathrm{time} = \mathrm{time} + \Delta t$
\EndWhile
\end{algorithmic}
\end{algorithm}

Using $S_{b,i} = 1 - S_{w,i}$ and $P_{b,i} = P_{w,i} + P_c(S_{b,i})$, the flow problem (Eqs.\ref{eq:wet} and \ref{eq:bub_flow}) is expressed in terms of two primary unknowns per pore: $P_{w,i}$ and $S_{b,i}$. The flow rates $q_{w,ij}$ and $q_{b,ij}$ are computed from Eq.\ref{eq:flux_q} using the conductivities in Section~\ref{sec:ipnm_throat}. Applying backward Euler yields the nonlinear system:
\begin{equation}
\label{eq:flow_residual}
\mathbf{F}(\mathbf{x}^{n+1}) = \begin{bmatrix} V_i(S_{w,i}^{n+1} - S_{w,i}^n) - \delta t \displaystyle\sum_{j=1}^{z_i} q_{w,ij}^{n+1} \\[10pt] V_i(S_{b,i}^{n+1} - S_{b,i}^n) - \delta t \displaystyle\sum_{j=1}^{z_i} q_{b,ij}^{n+1} \end{bmatrix} = \mathbf{0}
\end{equation}
where $\mathbf{x} = [P_{w,i};\, S_{b,i}]$ for all pores $i$. The nonlinearity is due to conductivities $K_w$ and $K_b$ (Eqs.\ref{eq:Kw} and \ref{eq:Kb}) and capillary pressure $P_c$ (Eqs.\ref{eq:kbsb_sng}--\ref{eq:kbsb_brg}) being nonlinear functions of $S_{b,i}$. We solve Eq.\ref{eq:flow_residual} via Newton's method with the Jacobian computed by automatic differentiation. Iterations terminate when $\|\mathbf{F}\|$, $\|\mathbf{F}\|/\|\mathbf{F}^0\|$, or $\|\Delta\mathbf{x}\|/\|\mathbf{x}\|$ falls below a tolerance. After each sub-step, a capillary stability check (Algorithm~\ref{alg:capequil}) is performed. Algorithm~\ref{alg:flow} summarizes the flow solver.

\begin{algorithm}[t]
\caption{Flow solver}\label{alg:flow}
\begin{algorithmic}[0]
\Statex \textbf{Input:} $P_{w,i}$, $S_{b,i}$, $\chi_i$, $\chi_{ij}$, $\Delta t$
\Statex \textbf{Output:} $P_{w,i}$, $S_{b,i}$, $\chi_i$, $\chi_{ij}$, $q_{w,ij}$
\vspace{2pt}
\State $\mathrm{time} = 0$;\; $\delta t = \Delta t$
\While{$\mathrm{time} < \Delta t$}
  \State $\mathbf{x}^{n,0} = [P_{w,i}^n;\, S_{b,i}^n]$
  \For{$k = 1, \ldots, k_{\max}$}
    \State Compute $\mathbf{F}(\mathbf{x}^{n,k})$ and $\mathbf{J}(\mathbf{x}^{n,k})$ from Eq.\ref{eq:flow_residual}
    \State $\Delta\mathbf{x} = -\mathbf{J}^{-1}\mathbf{F}$;\; $\mathbf{x}^{n,k+1} = \mathbf{x}^{n,k} + \Delta\mathbf{x}$
    \State If converged, exit
  \EndFor
  \State $[P_{w,i}^{n+1},\, S_{b,i}^{n+1}] = \mathbf{x}^{n,k+1}$
  \State Adapt $\delta t$
  \State $[S_{b,i},\, \chi_i,\, \chi_{ij}]$ = Capillary Stability ($S_{b,i}^{n+1}$, $\chi_i$, $\chi_{ij}$) \hfill [Algorithm~\ref{alg:capequil}]
  \State $\mathrm{time} = \mathrm{time} + \delta t$
\EndWhile
\State Compute $q_{w,ij}$ from $P_{w,i}$, $S_{b,i}$ via Eq.\ref{eq:flux_q}
\end{algorithmic}
\end{algorithm}

The solute flux $J_{ij}$ (Eq.\ref{eq:flux_J}) is a linear function of $C$, so the transport equations reduce to a linear system that is solved in a single step without Newton iteration. At each sub-step, the concentration in occupied pores is first set from local thermodynamic equilibrium (Eq.\ref{eq:henry}). These values serve as internal boundary conditions for the vacant pores, following the approach of \cite{mehmani2022AWR}. Applying backward Euler to Eq.\ref{eq:vac} for each vacant pore $i$ gives:
\begin{equation}
\label{eq:transport_vac}
V_i(C_i^{n+1} - C_i^n) = \delta t \sum_{j=1}^{z_i^w} J_{ij}^{n+1}
\end{equation}
The saturation in occupied pores is updated explicitly from Eq.\ref{eq:occ} using the concentration field at time level $n$:
\begin{equation}
\label{eq:transport_occ}
S_{b,i}^{n+1} = S_{b,i}^n + \frac{\rho_w}{\rho_b V_i} \sum_{j=1}^{z_i^w} J_{ij}^{n}\,\delta t
\end{equation}
After each sub-step, a capillary stability check (Algorithm~\ref{alg:capequil}) is performed. Algorithm~\ref{alg:transport} outlines the transport solver.

\begin{algorithm}[t]
\caption{Transport solver}\label{alg:transport}
\begin{algorithmic}[0]
\Statex \textbf{Input:} $C_i$, $S_{b,i}$, $q_{w,ij}$, $\chi_i$, $\chi_{ij}$, $\Delta t$
\Statex \textbf{Output:} $C_i$, $S_{b,i}$, $\chi_i$, $\chi_{ij}$
\vspace{2pt}
\State $\mathrm{time} = 0$;\; $\delta t = \Delta t$
\While{$\mathrm{time} < \Delta t$}
  \State Set $C_i^{n+1}$ from Eq.\ref{eq:henry} for occupied pores
  \State Compute $(dS_{b,i}/dt)^n$ from Eq.\ref{eq:transport_occ}
  \State Adapt $\delta t$
  \State Solve Eq.\ref{eq:transport_vac} for $C_i^{n+1}$ at vacant pores
  \State Update $S_{b,i}^{n+1}$ at occupied pores from Eq.\ref{eq:transport_occ}
  \State $[S_{b,i},\, \chi_i,\, \chi_{ij}]$ = Capillary Stability ($S_{b,i}^{n+1}$, $\chi_i$, $\chi_{ij}$) \hfill [Algorithm~\ref{alg:capequil}]
  \State $\mathrm{time} = \mathrm{time} + \delta t$
\EndWhile
\end{algorithmic}
\end{algorithm}

The flow solver (Algorithm~\ref{alg:flow}) adapts $\delta t$ based on the bubble saturation $S_{b,i}$ in each occupied pore (terminal or bridge). For each pore, the nearest \textit{event saturation} $S_{b,i}^x$ in the direction of $S_{b,i}$ change is identified. Events include the pore filling to its capacity $\smash{S_{b,i}^{\max}}$ (computed as the saturation at which $P_{c,i} \!=\! 100\,\max(P_{ce,ij})$, where $P_{ce,ij}$ are the entry pressures of the pore's unoccupied throats), the pore emptying ($S_{b,i} \!\to\! -10^{-6}$; the slightly negative bound ensures retraction is triggered in Algorithm~\ref{alg:capequil} restoring $S_{b,i} = 0$), and crossings of critical saturation ($S_{b,i} = S_c$), invasion, and snap-off thresholds.
The value of $S_{b,i}^{\max}$ is chosen large enough that invasion into adjacent throats occurs well before $S_{b,i}$ reaches it, while staying far from $S_{b,i} \!=\! 1$ where the $\kappa_b$--$S_b$ curve has a vertical asymptote.
End-point saturations $\smash{S_{b,i}^{\max}}$ and $-10^{-6}$ are hard limits, and the sub-step is rejected if these limits are violated. The intermediate $\smash{S_{b,i}^x}$ are allowed to overshoot by up to $10^{-3}$, ensuring that the solver can cross them rather than asymptotically approaching them. If Newton fails to converge, the step is rejected, $\delta t$ is halved, and the iterations are restarted. The next sub-step is estimated by the linear extrapolation $\delta t_{\mathrm{new}} = (|S_{b,i}^x - S_{b,i}^n|\,/\,|S_{b,i}^{n+1} - S_{b,i}^n|)\,\delta t$, followed by taking the minimum over all occupied pores. While we check for all event saturations in the simulations of Section~\ref{sec:exp}, we have found empirically that checking for only the end-points is sufficient and accelerates computations with negligible impact on accuracy.

The transport solver (Algorithm~\ref{alg:transport}) adapts $\delta t$ somewhat differently, by controlling the saturation increment per sub-step. Concretely, a trial increment $\Delta S_{b,i} = (dS_{b,i}/dt)^n\,\delta t$ is first computed for each occupied pore, then bounded to lie within prescribed limits $[\Delta S_{b,\min},\,\Delta S_{b,\max}]$ (here $[10^{-3},\, 10^{-1}]$). If the trial updated saturation $\smash{S_{b,i}^* = S_{b,i}^n + \Delta S_{b,i}}$ falls outside the interval $[-10^{-6},\, S_{b,\max}]$, the increment is further reduced. The sub-step is then back-calculated from $\delta t = \min_i(\Delta S_{b,i}\,/\,(dS_{b,i}/dt)^n)$ over all occupied pores that exchange solute through at least one unoccupied throat (Eq.\ref{eq:occ}). The outer coupling loop (Algorithm~\ref{alg:ipnm}) employs the same adaptive protocol. If either returns $\delta t = 0$, indicating stagnation, the outer time step $\Delta t$ is halved and both the flow and transport solvers are restarted.

\subsection{Visualization}
\label{sec:ipnm_vis}
\begin{figure}[t!]
\centering
\includegraphics[width=0.9\textwidth]{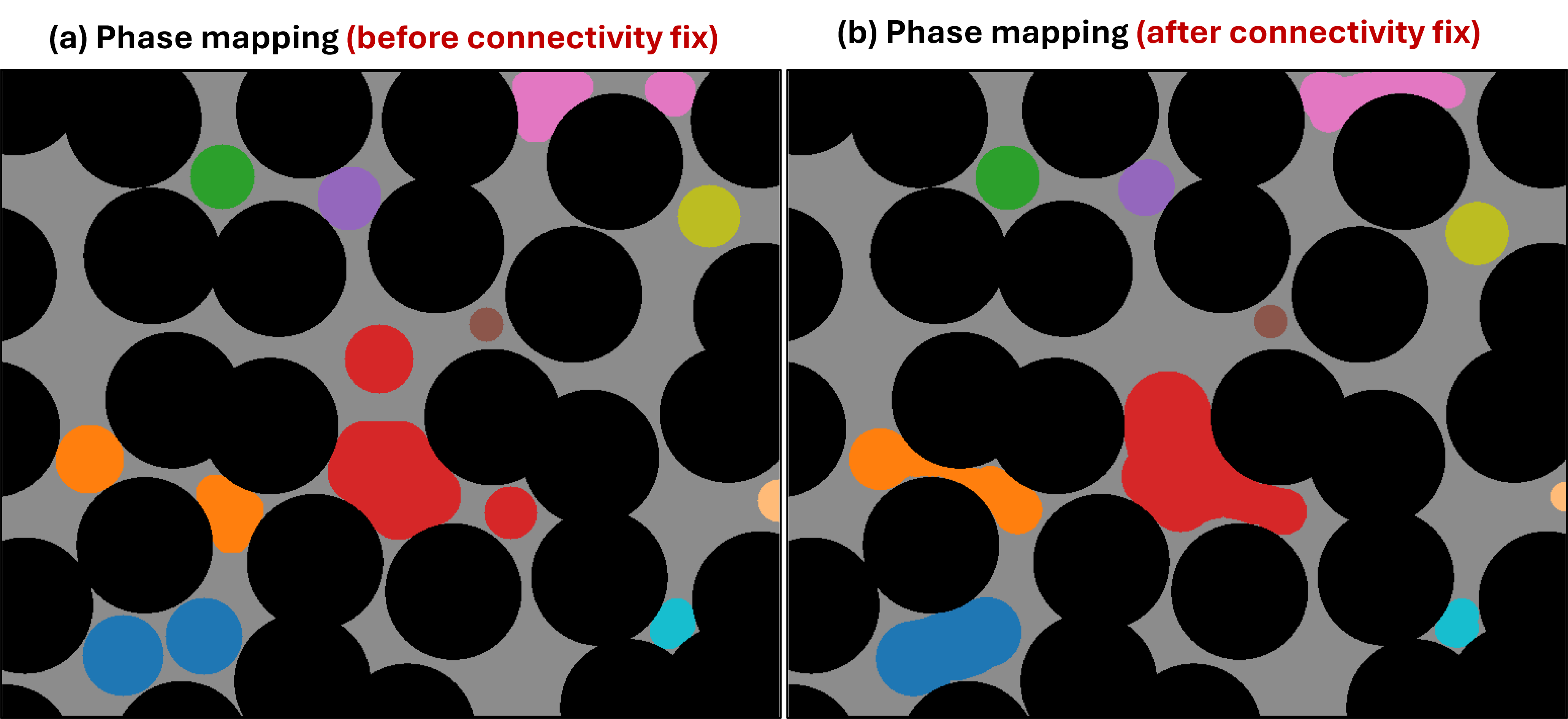}
\caption{Visualization of the non-wetting phase distribution mapped onto the original pore-space image. Each ganglion is assigned a unique color. (a) Per-pore mapping using PMM subimages: ganglia spanning multiple pores appear as disconnected fragments (e.g., the red ganglion occupying three adjacent pores). (b) Connectivity-corrected mapping: the medial axis (skeleton) of the pore space is used to bridge fragments across pores belonging to the same ganglion, restoring physical connectivity while preserving the local curvature and total ganglion volume.}
\label{fig:visual}
\end{figure}

A key feature of iPNM is that the bubble saturations $S_{b,i}$ computed for each pore can be mapped directly onto the original pore-space image. For each occupied pore, the PMM subimage corresponding to $S_{b,i}$ provides the spatial distribution of the non-wetting phase within that pore (Section~\ref{sec:ipnm_pore}). Because the $\kappa_b$--$S_b$ curves are constructed from the actual pore geometry, this mapping is faithful to the underlying microstructure. This is in contrast to traditional PNMs, which approximate pore shapes with idealized geometries (e.g., spheres, cubes) and cannot reconstruct a high-fidelity phase distribution. A naive per-pore mapping, however, produces disconnected fragments for ganglia that span multiple pores (Fig.\ref{fig:visual}a; see red ganglion occupying three adjacent pores). To restore visual connectivity, we first identify which pores belong to the same ganglion via connected-component labeling of the subgraph defined by $\chi_i$ and $\chi_{ij}$ (Section~\ref{sec:ipnm_cap}). Each ganglion is assigned a unique color (Fig.\ref{fig:visual}). The medial axis (skeleton) of the pore space is then used to bridge fragments across adjacent pores within each ganglion, preserving local curvature and matching the ganglion's total volume (Fig.\ref{fig:visual}b).
Such a connectivity-corrected visualization is used in Section~\ref{sec:exp}.
While iPNM does not impose uniform $P_c$ within a ganglion, the balance equations naturally drive it toward uniformity, consistent with experimental evidence~\cite{salehpour2025micro}. Transient deviations arise during capillary events, which induce intra-ganglion flow.

\section{Numerical verification}
\label{sec:num}
In this section, we verify that the proposed iPNM accurately reproduces key pore-scale phenomena associated with Ostwald ripening, before deploying it to complex domains in Section~\ref{sec:exp}.
The verification cases were originally presented in~\cite{mehmani2022AWR} to validate a quasi-static PNM against microfluidic experiments and direct numerical simulations. That PNM enforces instantaneous capillary equilibration within each ganglion at all times, whereas iPNM resolves the finite timescale of this process through its two-phase flow solver. Reproducing the quasi-static predictions ensures that iPNM retains the validated behavior.
In all cases, pores are semi-cubic: a pore with inscribed-sphere radius $R_p$ and volume $(2R_p)^3$. The pore shape is controlled by fitting parameters $\lambda_1$ and $\lambda_2$ in Eqs.\ref{eq:kbsb_sng}--\ref{eq:kbsb_brg}, which we set to $\lambda_1 = 0.905$, $\lambda_2 = 0.01$, and $n = 2$, matching those in~\cite{mehmani2022AWR}. Throats are prismatic with square cross-sections, with side half-length $R_t = \xi \times \min(R_{p1}, R_{p2})$, where $\xi$ is the throat-to-pore aspect ratio, and length $L_t = L_s - R_{p1} - R_{p2}$, where $L_s$ is the lattice spacing. Networks have a 2D lattice topology but non-uniform out-of-plane thickness, since each pore's depth equals $2R_p$. Further details on the network geometry and boundary conditions can be found in~\cite{mehmani2022AWR}.
Below, we compare iPNM against the quasi-static PNM~\cite{mehmani2022AWR} for three verification cases: dissolution-induced snap-off (Section~\ref{sec:num_dissol}), bubble dislocation (Section~\ref{sec:num_disloc}), and ganglion growth (Section~\ref{sec:num_gang}).

\subsection{Dissolution-induced snap-off}
\label{sec:num_dissol}
\begin{figure}[t!]
\centering
\includegraphics[width=0.9\textwidth, trim=15 10 13 15, clip]{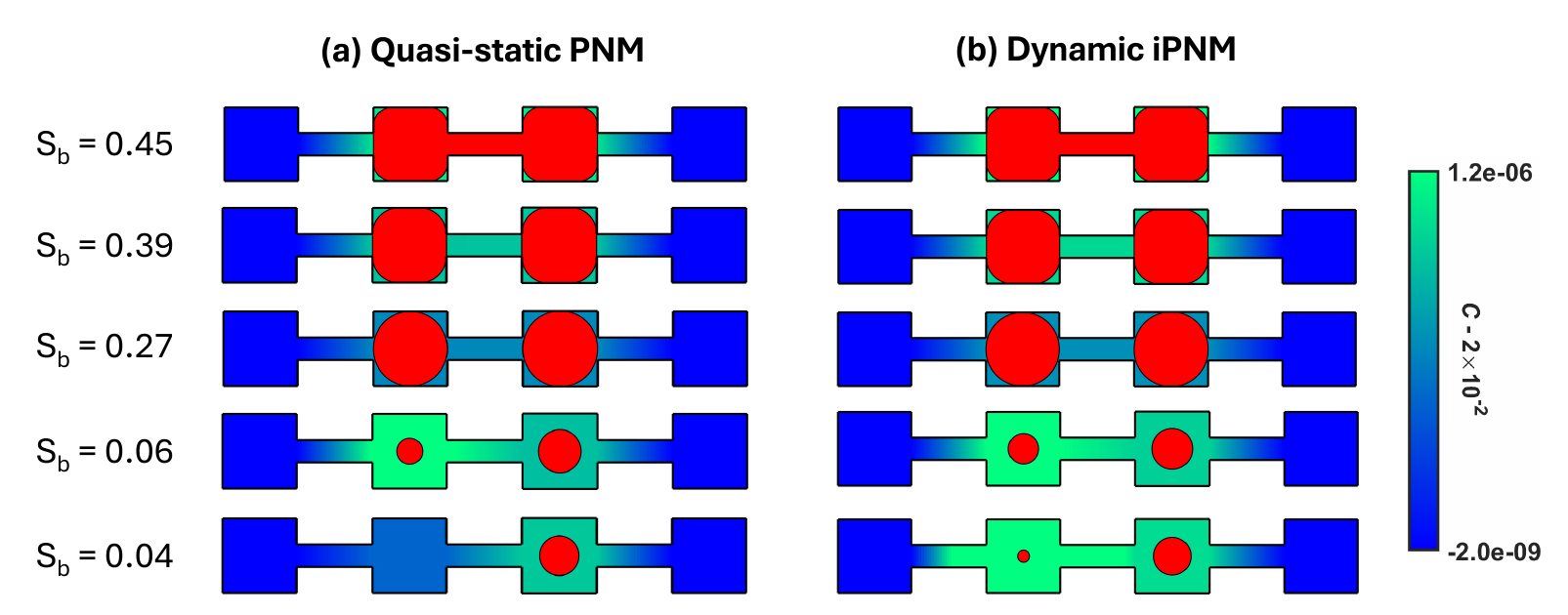}
\caption{Comparison of dissolution-induced snap-off and ripening between 
(a) quasi-static PNM~\cite{mehmani2022AWR} and (b) iPNM. Snapshots show 
the non-wetting phase distribution (red) and dissolved gas concentration 
field (color map) at decreasing bubble saturations (annotated on the left). Snap-off occurs at $S_b \approx 0.39$, followed by dissolution and ripening of the ganglion fragments. Throats carry no volume in either model.}
\label{fig:dissol}
\end{figure}

We first examine dissolution-induced snap-off, observed in a microfluidic experiment by Sahloul et al.~\cite{sahloul2002NAPL}, in which a ganglion spanning two pores dissolves into the surrounding wetting phase. The ganglion fragments into two singleton bubbles that then undergo simultaneous dissolution and ripening.
This sequence was successfully captured by the quasi-static PNM of \cite{mehmani2022AWR}, which we now reproduce via iPNM.
Following \cite{mehmani2022AWR}, we consider a four-pore network in which a ganglion initially occupies the two central pores and the connecting throat (Fig.\ref{fig:dissol}a, $S_b = 0.45$). A small perturbation in pore size (${\sim}1\%$) is introduced to break symmetry. Dissolution is driven by imposing a lower dissolved concentration at the two outer boundary pores, which act as concentration sinks similar to the experiment. A constant wetting-phase pressure is imposed at both boundaries to allow the wetting phase to enter or leave in response to volumetric changes of the bubbles. As shown in Fig.\ref{fig:dissol}, both models predict snap-off of the ganglion into two singleton bubbles at $S_b \approx 0.39$, followed by continued dissolution and ripening of the fragments. In the quasi-static PNM, capillary equilibrium within the connected ganglion prior to snap-off is enforced instantaneously. In iPNM, this equilibration proceeds dynamically through two-phase flow. As the non-wetting phase redistributes between the two occupied pores, the wetting phase adjusts to accommodate the resulting volume changes. This finite equilibration timescale leads to minor differences in the spatial distribution at later stages ($S_b = 0.06$ and $S_b = 0.04$).

\subsection{Bubble dislocation}
\label{sec:num_disloc}
\begin{figure}[t!]
\centering
\includegraphics[width=0.95\textwidth, trim=15 10 10 15, clip]{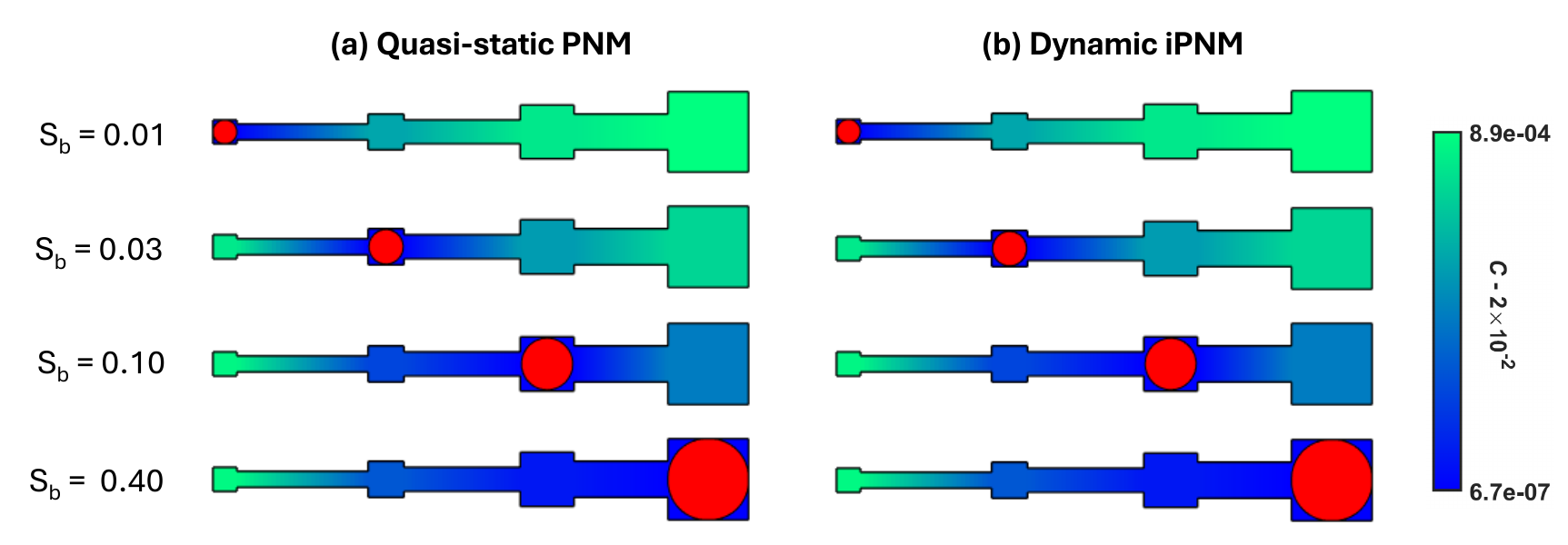}
\caption{Comparison of bubble dislocation between (a) quasi-static 
PNM~\cite{mehmani2022AWR} and (b) iPNM. Snapshots show the non-wetting 
phase distribution (red) and dissolved gas concentration field (color map) at increasing bubble saturations (annotated on the left). The bubble migrates sequentially from the smallest to the largest pore in a 
four-pore network. Throats carry no volume in either model.}
\label{fig:disloc}
\end{figure}

Mehmani and Xu~\cite{mehmani2022AWR} coined \textit{bubble dislocation} to describe the pore-to-pore migration of a growing bubble from a small pore into an adjacent large pore. The mechanism arises because, upon invading the large pore, the bubble lacks sufficient volume to be halted by that pore's walls. Hence, the bubble retracts from the small pore and moves entirely into the large pore, which is energetically favorable. This phenomenon was demonstrated by DNS and observed in microfluidic experiments. The quasi-static PNM of \cite{mehmani2022AWR} captured this behavior, which we now reproduce with iPNM. Following~\cite{mehmani2022AWR}, we consider a four-pore network with pores arranged in ascending size from left to right. A bubble is initially placed in the smallest pore (Fig.\ref{fig:disloc}a, $S_b = 0.01$). Slightly elevated dissolved concentrations are imposed at boundary pores (not depicted) connected to each end of the network, driving growth via diffusive mass uptake. A constant wetting-phase pressure is imposed at both boundaries to allow the wetting phase to enter or leave. In the quasi-static PNM, the bubble jumps to the next pore at the moment of invasion. In iPNM, this transition has a finite timescale, because the non-wetting phase must flow into the larger pore while the wetting phase flows countercurrent through corner films. Fig.\ref{fig:disloc}b shows iPNM reproduces the same sequence of events, with the bubble migrating through progressively larger pores. We note dislocation is challenging to model because it involves two events (invasion and retraction) occurring simultaneously, which taxes the time stepping and Newton iterations of the flow solver.

\subsection{Ganglion growth}
\label{sec:num_gang}
We next simulate ganglion growth in $11\times11$ heterogeneous pore networks for two aspect ratios, $\xi = 1/4$ and $\xi = 3/4$, following the supplementary material of~\cite{mehmani2022AWR}. Pore sizes are drawn from a uniform distribution with contrast ratio $R_{p,\max}/R_{p,\min} = 2$, where $R_{p,\max}$ and $R_{p,\min}$ are the maximum inscribed-sphere radii of the largest and smallest pores. Boundary throats are made intentionally small to create a capillary barrier that prevents the non-wetting phase from leaving the network. A small bubble is initially placed at the center of the network, and elevated dissolved concentrations are imposed at all boundary pores to drive growth. A constant wetting-phase pressure is imposed at the boundaries, as in Sections~\ref{sec:num_dissol}--\ref{sec:num_disloc}. Simulations continue until $S_b^{tot}$ reaches 95\%. As the ganglion grows, variations in pore and throat geometry cause it to repeatedly fragment and reconnect, producing a complex topological evolution.

\begin{figure}[t!]
\centering
\includegraphics[width=0.8\textwidth, trim=10 5 20 10, clip]{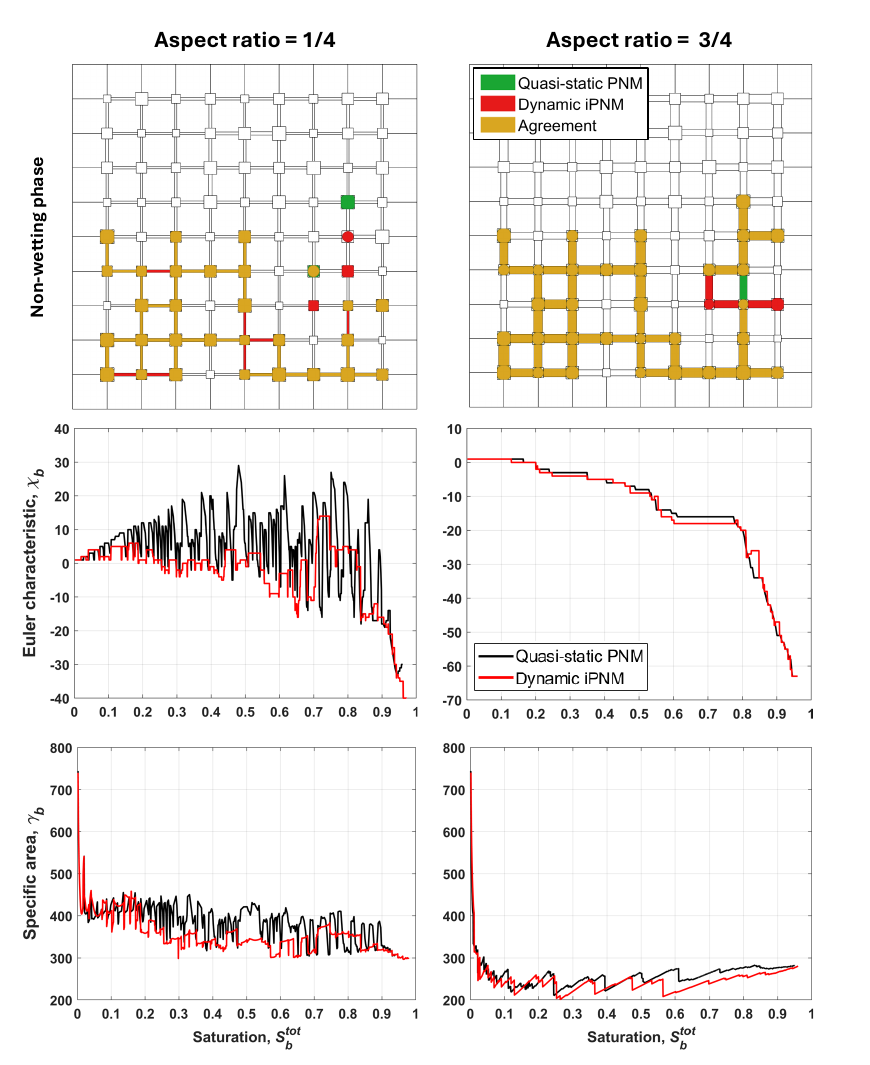}
\caption{Comparison of ganglion growth between quasi-static PNM~\cite{mehmani2022AWR} and iPNM for two aspect ratios: $\xi = 1/4$ (left column) and $\xi = 3/4$ (right column). Top row shows the non-wetting phase distribution (red: quasi-static PNM; green: iPNM). Middle and bottom rows show the Euler characteristic $\chi_b$ and specific interfacial area $\gamma_b$, respectively, as functions of total saturation $S_b^{tot}$ (black: quasi-static PNM; red: iPNM).}
\label{fig:grwth}
\end{figure}

We track two descriptors of the ganglion to compare iPNM against the quasi-static PNM: (1) Euler characteristic $\chi_b = N - L + O$, where $N$ is the number of disconnected ganglia, $L$ is the number of redundant loops, and $O$ is the number of internal voids (zero for the 2D lattice here). A high positive $\chi_b$ entails a fragmented phase, while a low or negative value entails increased connectivity; (2) Specific area $\gamma_b = \Gamma_b / V_b$, where $\Gamma_b$ is the total gas--liquid interfacial area and $V_b$ is the total non-wetting phase volume. Both PNMs are run on the same networks with these metrics compared in Fig.\ref{fig:grwth}. The models agree on the overall oscillatory behavior of $\chi_b$ and $\gamma_b$ as functions of $S_b^{tot}$. Agreement is closer for $\xi = 3/4$, where growth is continuous and snap-off is suppressed. For $\xi = 1/4$, the ganglion fragments and reconnects frequently, so mispredicting a single event can snowball into larger pointwise deviations between the two models. These deviations stem from iPNM resolving the finite timescale of capillary equilibration, which the quasi-static PNM does not. Nevertheless, the overall statistical behavior remains in agreement.

\section{Experimental validation}
\label{sec:exp}

\begin{figure}[b!]
\centering
\includegraphics[width=1.0\textwidth, trim=13 10 10 3, clip]{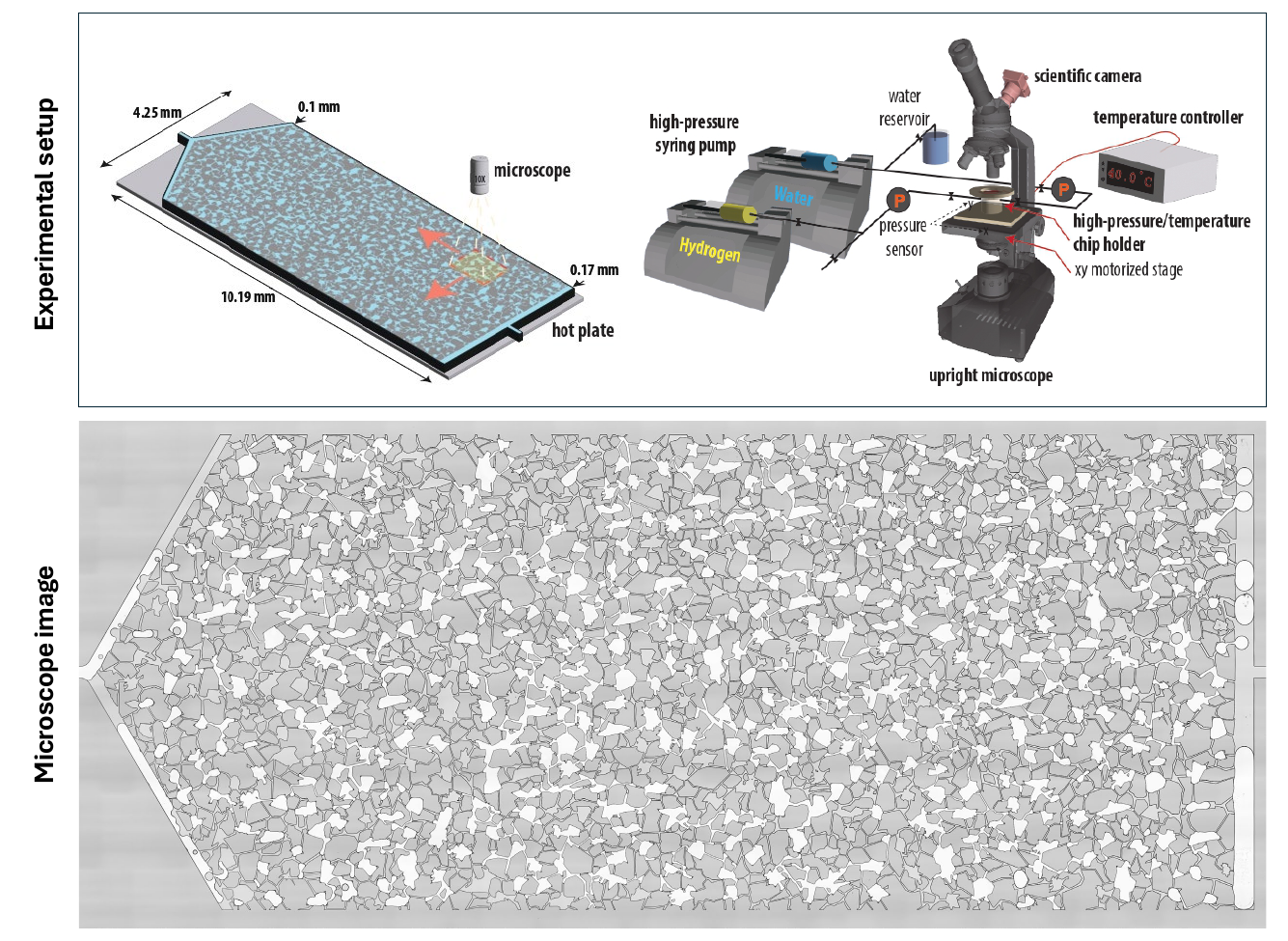}
\caption{Experimental setup (top) showing the silicon-glass bonded micromodel, high-pressure syringe pump, and temperature controller. Raw microscope image (bottom) of the micromodel showing residually trapped hydrogen bubbles (white) within the sandstone pore space (gray) at $t = 1$~hr in the 40HS case. The wider inlet and outlet channels on the left and right edges contain large bubbles that act as concentration sinks.}
\label{fig:exp}
\end{figure}

We next benchmark iPNM against the microfluidic experiments of Salehpour et al.~\cite{salehpour2025micro}, which track the Ostwald ripening of residually trapped hydrogen ganglia in a silicon-glass bonded micromodel patterned from a 2D X-ray CT image of a Canadian sandstone. The micromodel measures $10.19 \times 4.25$~mm in the horizontal plane with a uniform gap thickness of $g \!=\! 7.8$~\textmu m, and has a median pore size of 42~\textmu m, a median throat size of 10.6~\textmu m, and a porosity of 0.3. These dimensions introduce sufficient pore-scale heterogeneity for the iPNM to demonstrate its capability in handling realistic pore geometries. For a detailed description of the experimental apparatus and imaging protocol, see~\cite{salehpour2025micro}. A schematic of the experimental setup and a representative raw image of the micromodel are shown in Fig.\ref{fig:exp}.

Experiments were conducted at two temperatures (40\textdegree C and 80\textdegree C). Hydrogen was first injected into the water-saturated micromodel, followed by water reinjection to residually trap the gas as disconnected bubbles within the pore space. Large bubbles were deliberately retained in the wider inlet and outlet channels ($R_l = 50$~\textmu m and $R_r = 85$~\textmu m, respectively) so that they served as concentration sinks, sustaining diffusive mass transfer from the porous matrix toward the channels. Four experimental cases are considered, combining two temperatures with two levels of initial gas saturation: 40\textdegree C low-saturation (40LS), 40\textdegree C high-saturation (40HS), 80\textdegree C low-saturation (80LS), and 80\textdegree C high-saturation (80HS). The initial gas saturations range from 0.47 to 0.67, and experiments lasted 15--24 days.\footnote{The experimental images were reprocessed with improved algorithms, yielding saturations and bubble statistics that differ slightly from~\cite{salehpour2025micro}.}

We compare iPNM against both the experimental data and the continuum model of~\cite{salehpour2025micro}. The continuum model predicts the spatiotemporal evolution of dissolved concentration and gas saturation along the longitudinal direction of the micromodel. It couples Henry's law with a $P_c$--$S_b$ relationship obtained by applying PMM to the entire micromodel image. Unlike iPNM, the continuum model assumes local capillary equilibrium within each representative elementary volume (REV) and therefore cannot resolve individual bubble curvatures or population statistics. The model equations and numerical implementation are given in \ref{app:continuum}. In what follows, we first describe the iPNM simulation setup (Section~\ref{sec:ipnm_setup}), then compare macroscopic quantities (Section~\ref{sec:macro}) and population statistics (Section~\ref{sec:popstat}).

\subsection{iPNM simulation setup}
\label{sec:ipnm_setup}

The pore network is extracted from the binary image of the micromodel via marker-based watershed segmentation (Section~\ref{sec:probdesc}). For each pore $i$, the volume $V_i$ and maximum inscribed radius $R_{p,i}$ are computed, and PMM is applied pore-wise to construct the $\kappa_b$--$S_b$ curve as described in Section~\ref{sec:ipnm_pore}. Throats are modeled as rectangular cross-sections with in-plane half-width $a$ equal to the inscribed radius at the constriction, out-of-plane half-width $b = g/2$, and length $L_t$ equal to the sum of Euclidean distances from the constriction to the centroids of its two neighboring pores.

The initial saturation field is obtained by mapping bubble volumes segmented from the experimental image at $t = 0$ onto the extracted pore network. Dirichlet BCs on $C$ are imposed at the left and right boundary throats by evaluating Henry's law (Eq.~\ref{eq:henry}) with the capillary pressure of an interface spanning each channel's half-width, $P_c = \sigma(1/R_{l,r} + 2/g)$, consistent with \ref{app:continuum}. The remaining boundaries are treated as no-flux. For the flow problem, a constant wetting-phase pressure $P_w = 1$~atm is imposed at the left and right boundaries, with no-flow on the remaining sides, so that gas redistribution is driven entirely by diffusive mass transfer. The non-wetting phase molar density is computed from the ideal gas law as $\rho_b = P_w/(R_g T)$ and held constant throughout. This is because capillary pressure variations are small relative to the ambient pressure: $P_c/P_w \approx 3\%$ for the median pore size here. The water viscosity $\mu_w$ is computed from the correlation of Reid et al.~\cite{reid1987}, and the hydrogen viscosity $\mu_b$ is linearly interpolated between reference values at 300~K and 400~K from Mehl et al.~\cite{mehl2010}. All fluid properties are listed in Table~\ref{tab:fluid}.

\begin{table}[h!]
\centering
\caption{Fluid properties for each experimental temperature.}
\label{tab:fluid}
\setlength{\tabcolsep}{12pt}
\renewcommand{\arraystretch}{1.3}
\begin{tabular}{llccc}
\hline
Property & Symbol & 40\textdegree C & 80\textdegree C & Unit \\
\hline
Molecular diffusivity & $D_m$    & $7.34\times10^{-5}$ & $1.53\times10^{-4}$ & cm$^2$/s \\
Surface tension       & $\sigma$ & $68.9$              & $62.6$              & dyn/cm \\
Henry's constant      & $H$      & $7.51\times10^{10}$ & $7.55\times10^{10}$ & dyn/cm$^2$ \\
Vapor pressure        & $P_v$    & $7.36\times10^{4}$  & $4.73\times10^{5}$  & dyn/cm$^2$ \\
Water viscosity       & $\mu_w$  & $6.70\times10^{-3}$ & $3.60\times10^{-3}$ & poise \\
H$_2$ viscosity       & $\mu_b$  & $9.16\times10^{-5}$ & $9.96\times10^{-5}$ & poise \\
Water molar density   & $\rho_w$ & \multicolumn{2}{c}{$0.055$}             & mol/cm$^3$ \\
H$_2$ molar density   & $\rho_b$ & $3.89\times10^{-5}$ & $3.45\times10^{-5}$ & mol/cm$^3$ \\
\hline
\end{tabular}
\end{table}

\subsection{Macroscopic behavior}
\label{sec:macro}

\begin{figure}[t!]
\centering
\includegraphics[width=0.9\textwidth, trim=10 10 13 13, clip]{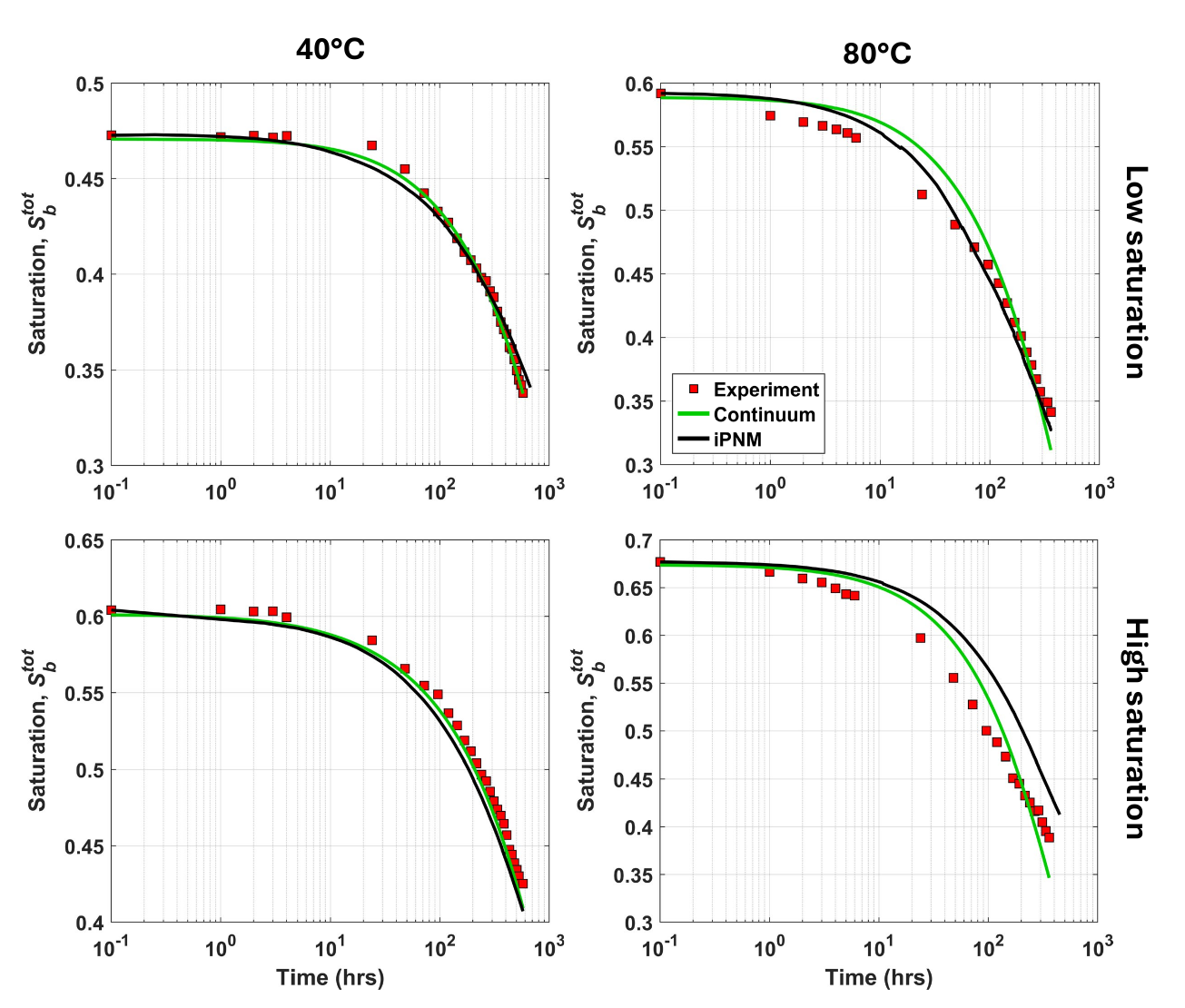}
\caption{Temporal evolution of total non-wetting phase saturation $S_b^{tot}$ for all four experimental cases. Red squares denote experiments, the black line is the iPNM prediction, and the green line is the continuum model. Cases correspond to 40\textdegree C and 80\textdegree C with low and high initial $S_b^{tot}$.}
\label{fig:sat}
\end{figure}

\noindent\textbf{Total saturation.}
We compare the temporal evolution of the total non-wetting phase saturation $S_b^{tot}$ in the micromodel across all four experimental cases (Fig.\ref{fig:sat}). In all cases, $S_b^{tot}$ declines monotonically, driven by diffusive transport toward the boundary channels. Both iPNM and the continuum model capture this trend. iPNM shows strong agreement with experiments, except for some deviation at later times in the 80HS case.
The deviation is attributed primarily to non-ideal BCs at the channels. In iPNM, the concentration at each channel is set uniformly based on the capillary pressure of an interface spanning the channel's half-width. In the experiments, however, the bubble distribution in the channels is highly irregular (Fig.\ref{fig:exp}), and in some cases low-curvature interfaces form \textit{along} (rather than \textit{across}) the channel, creating locally stronger sinks than assumed (see supplementary videos).
A likely secondary contributor is the presence of condensation droplets within trapped ganglia, which are most prevalent in the 80HS case (yellow in Fig.\ref{fig:spt80HS}). These droplets often condense and re-evaporate cyclically, with non-trivial effects on concentration and capillary pressure fields that are not captured by iPNM. Both sources of error are discussed further in Section~\ref{sec:disc_exp}.

\begin{figure}[t!]
\centering
\includegraphics[width=0.9\textwidth, trim=5 5 11 11, clip]{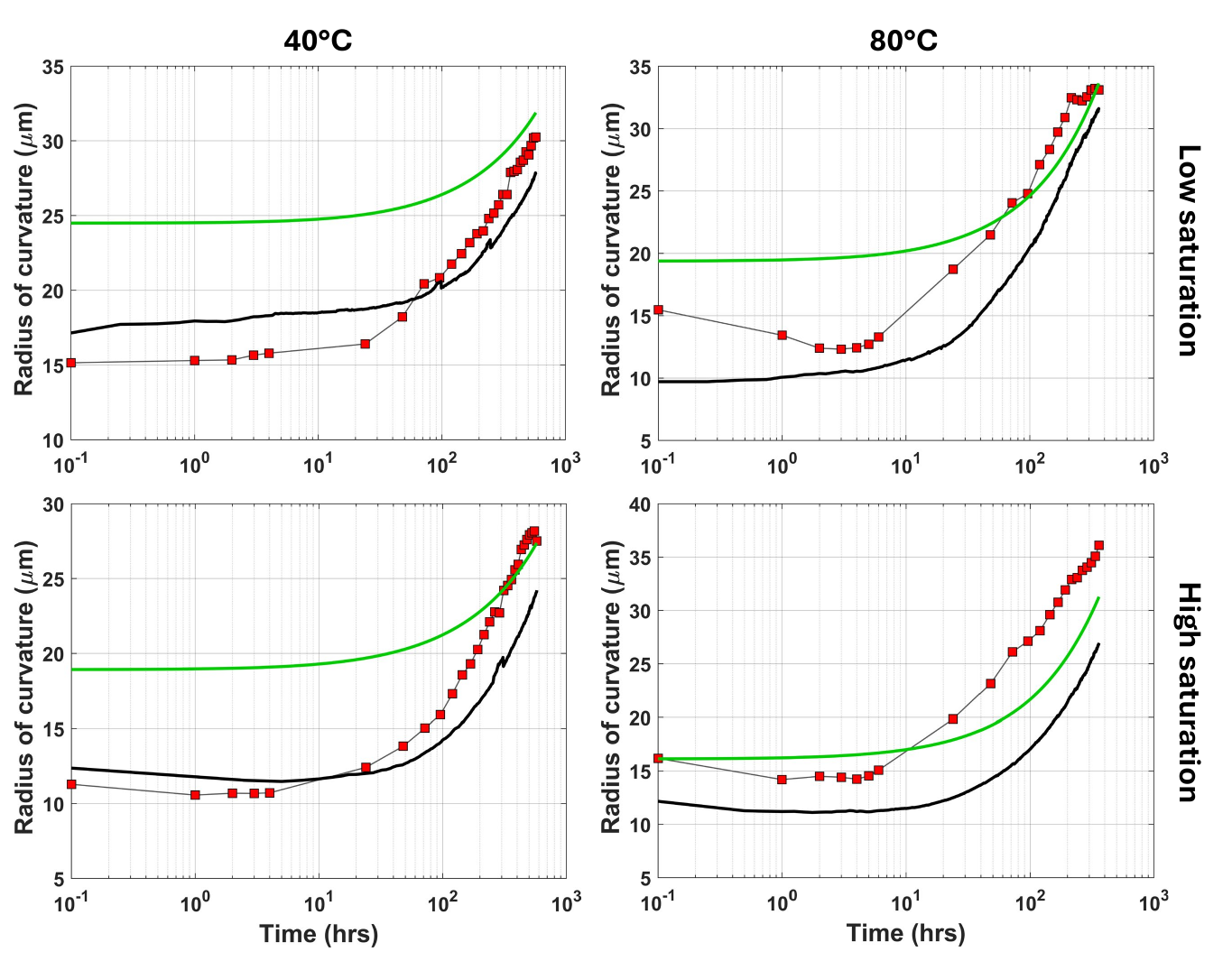}
\caption{Temporal evolution of the mean in-plane radius of curvature $\bar{r}$ for all four experimental cases. Red squares denote experiments, the black line is the iPNM prediction, and the green line is the continuum model. Cases correspond to 40\textdegree C and 80\textdegree C with low and high initial $S_b^{tot}$.}
\label{fig:curv}
\end{figure}

\medskip
\noindent\textbf{Mean curvature.}
Next, we compare the temporal evolution of the mean in-plane radius of curvature $\bar{r}$ in Fig.\ref{fig:curv}. In experiments, $\bar{r}$ is calculated by fitting circles to individual bubble interfaces extracted from acquired images using least-squares minimization, with quality control to exclude nearly straight arcs and interfaces with insufficient pixel coverage. These per-pore radii are averaged across each ganglion, then across all ganglia. In iPNM, the per-pore radii are obtained by evaluating the $\kappa_b$--$S_b$ curves, subtracting the out-of-plane contribution $\kappa_{gap} = 2/g$, then computing the reciprocal of the resulting in-plane curvature. The per-pore radii are averaged across ganglia similar to the experiments. In the continuum model, $\bar{r}$ is derived by mapping $S_b^{tot}$ to a capillary pressure via the global $P_c$--$S_b$ curve, then subtracting the out-of-plane contribution, and taking the reciprocal of the resulting in-plane curvature.

In all cases, Fig.\ref{fig:curv} shows that $\bar{r}$ increases monotonically over time. iPNM yields close quantitative agreement with experiments across all cases, though it tends to slightly underestimate $\bar{r}$ (particularly at 80\textdegree C and at late times for 40\textdegree C). This is partly attributable to the PMM-based $\kappa_b$--$S_b$ curves, which overestimate curvature for multi-pore ganglia, as examined further in Section~\ref{sec:popstat}. By contrast, the continuum model overestimates $\bar{r}$ at early and intermediate times, but approaches experimental data at late times. This arises because individual ganglion curvatures are not resolved, and $S_b^{tot}$ is mapped to a single curvature value using the global $P_c$--$S_b$ curve. This misses the small, high-curvature, and out-of-equilibrium bubbles that can pull the true population mean $\bar{r}$ downward. As these bubbles vanish over time, the curvature distribution narrows and the continuum model approaches the experiment.

\subsection{Pore-scale behavior}
\label{sec:popstat}

The macroscopic quantities of the previous section mask pore-scale information about ganglia. Here, we validate iPNM against experiments by comparing probability distributions of ganglion curvatures and spatial configurations of the non-wetting phase, capillary pressure, and dissolved concentration. The continuum model is inherently incapable of providing ganglion-level information, so its mean-property predictions are juxtaposed where appropriate.

\begin{figure}[t!]
\centering
\includegraphics[width=\textwidth]{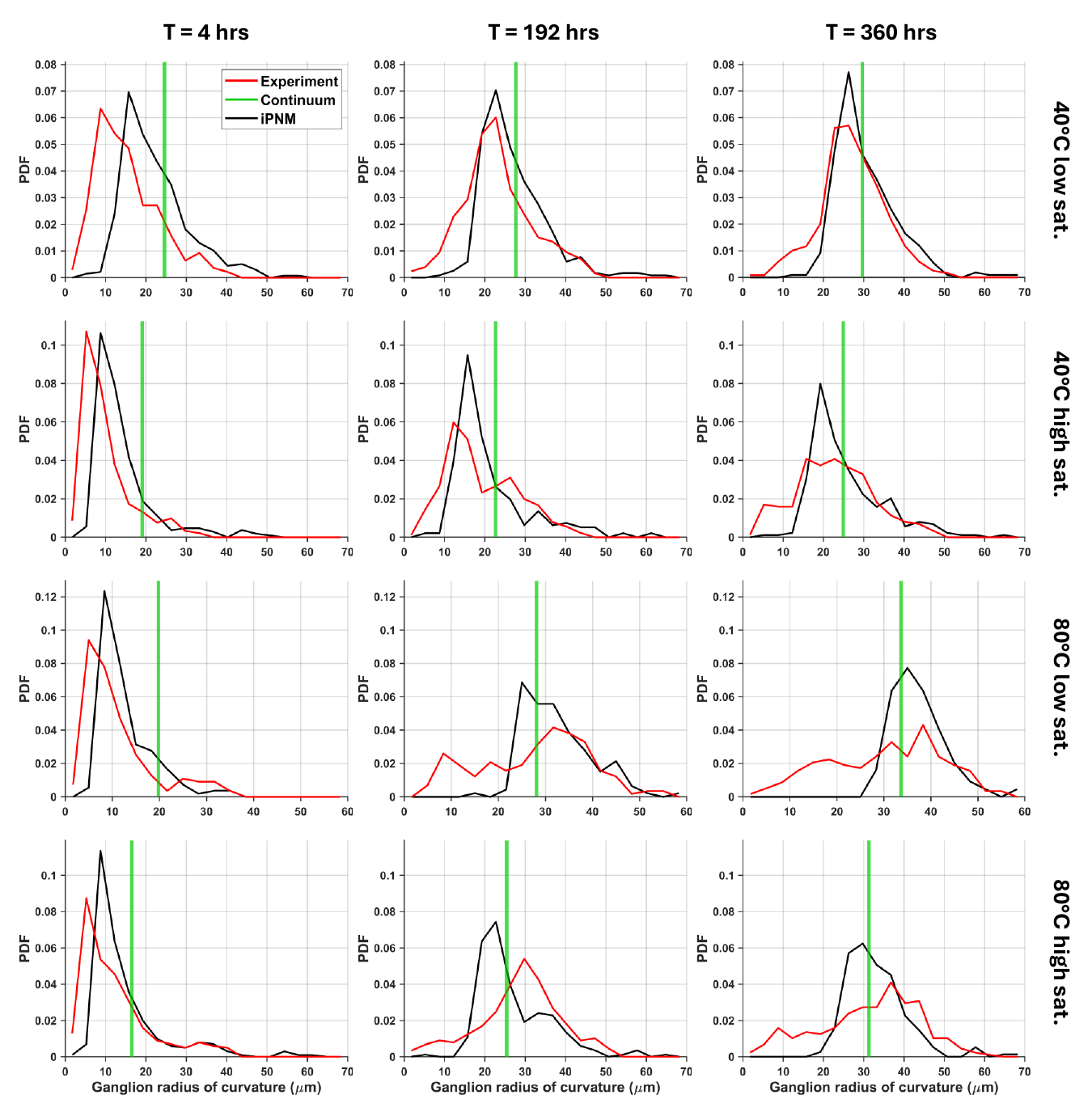}
\caption{PDF of the mean radius of curvature of ganglia at three times ($T = 4$, 192, and 360~hrs) for all four experimental cases. Red lines denote experimental data, black lines the iPNM predictions, and vertical green lines the continuum model's prediction of $\bar{r}$.}
\label{fig:curvPDF}
\end{figure}

\medskip
\noindent\textbf{Curvature distribution.}
Fig.\ref{fig:curvPDF} compares the PDF of radius of curvature of ganglia at three times ($T = 4$, 192, and 360~hrs). In all cases, we see that the experimental PDFs shift toward larger $r$ while simultaneously broadening. This reflects two concurrent processes governing ripening in the presence of boundary sinks~\cite{salehpour2025micro}: (1) \textit{Local} ripening, which is driven by the interaction between adjacent ganglia, forcing the PDFs to sharpen into a Dirac delta. Salehpour et al.~\cite{salehpour2025micro} showed local ripening lasts $\sim$5~hrs in the central region of the micromodel at 80\textdegree C, consistent with the $\sim$24~hr plateau of $\bar{r}$ in Fig.\ref{fig:curv} for the full domain; (2) \textit{Global} ripening, which is driven by diffusive mass loss to the channels. As most ganglia are supercritical (deformed), mass loss lowers curvature and shifts the PDF to larger $r$.

iPNM captures the peak positions and shifts of the PDFs across all cases. The agreement is strongest for 40\textdegree C, but deteriorates for 80\textdegree C at intermediate and late times. At 80\textdegree C, the experimental PDFs are broader than iPNM and have more pronounced tails toward small $r$.
We attribute this primarily to the irregular bubbles in the inlet and outlet channels (Fig.\ref{fig:exp}), whose effective curvatures---thus the boundary concentrations they impose---differ from the uniform values assumed in iPNM. A possible secondary factor is the higher prevalence of water condensation droplets at 80\textdegree C, whose cyclic condensation and re-evaporation alters concentration and capillary pressure fields in ways not captured by iPNM.
In Fig.\ref{fig:curvPDF}, the continuum model is shown as vertical lines denoting $\bar{r}$, which fall consistently to the right of experimental and iPNM peaks, except at late times.
This explains the overestimation of $\bar{r}$ in Fig.\ref{fig:curv} and is consistent with the assertion that the continuum model does not capture pre-equilibrium dynamics of ripening in an REV.

\begin{figure}[t!]
\centering
\hspace*{-1cm}
\includegraphics[width=1.1\textwidth]{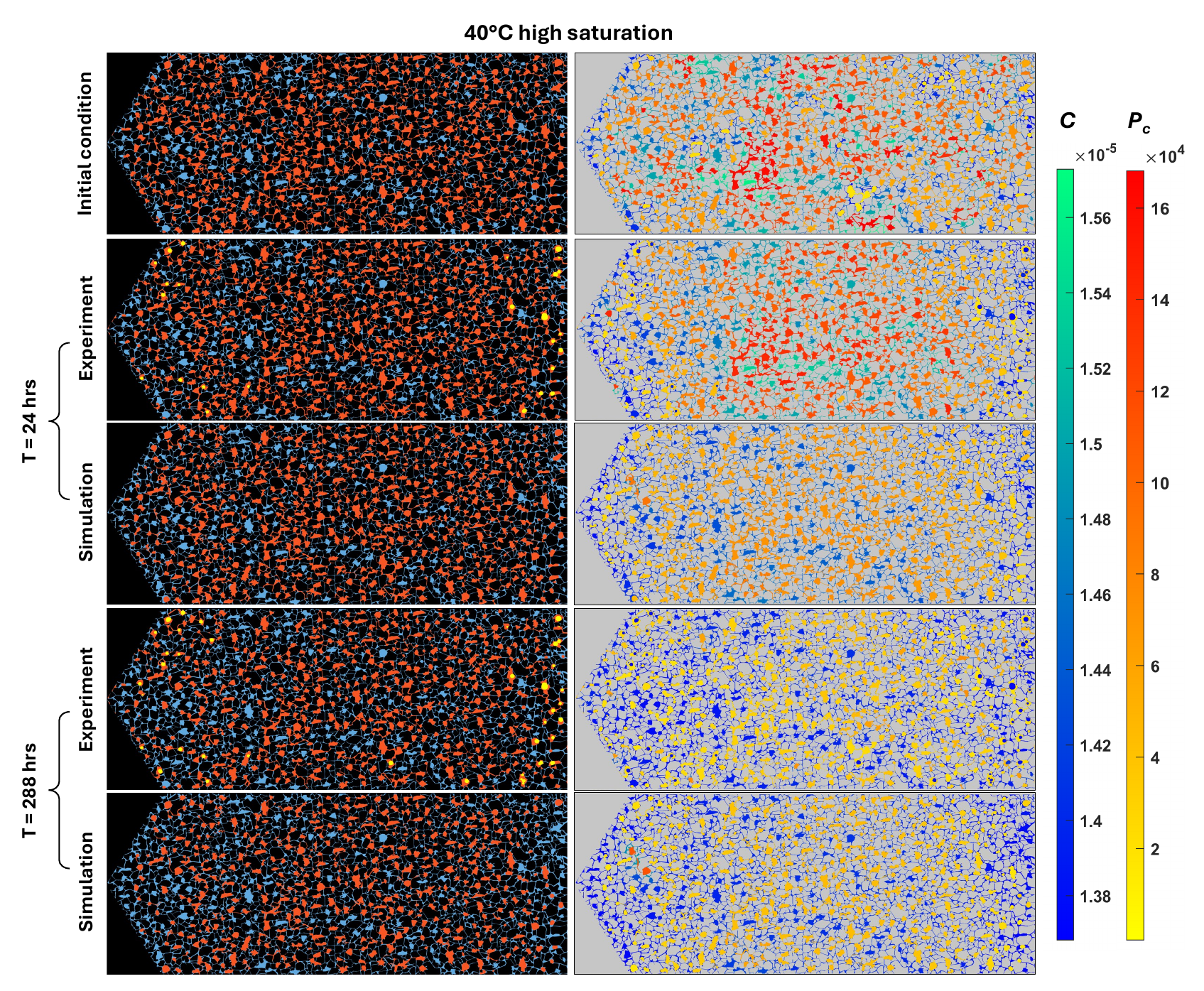}
\caption{Spatial distributions for the 40HS case. Left: phase distribution (red: non-wetting, blue: wetting, black: solid, yellow: condensation). Right: ganglia colored by capillary pressure (dyn/cm$^2$) and wetting phase by dissolved concentration (mole fraction). The top row shows the experimental initial condition used to initialize the simulation. Subsequent rows compare experiments and iPNM at $T \!=\! 24$ and $288$~hrs.}
\label{fig:spt40HS}
\end{figure}

\begin{figure}[t!]
\centering
\hspace*{-1cm}
\includegraphics[width=1.1\textwidth]{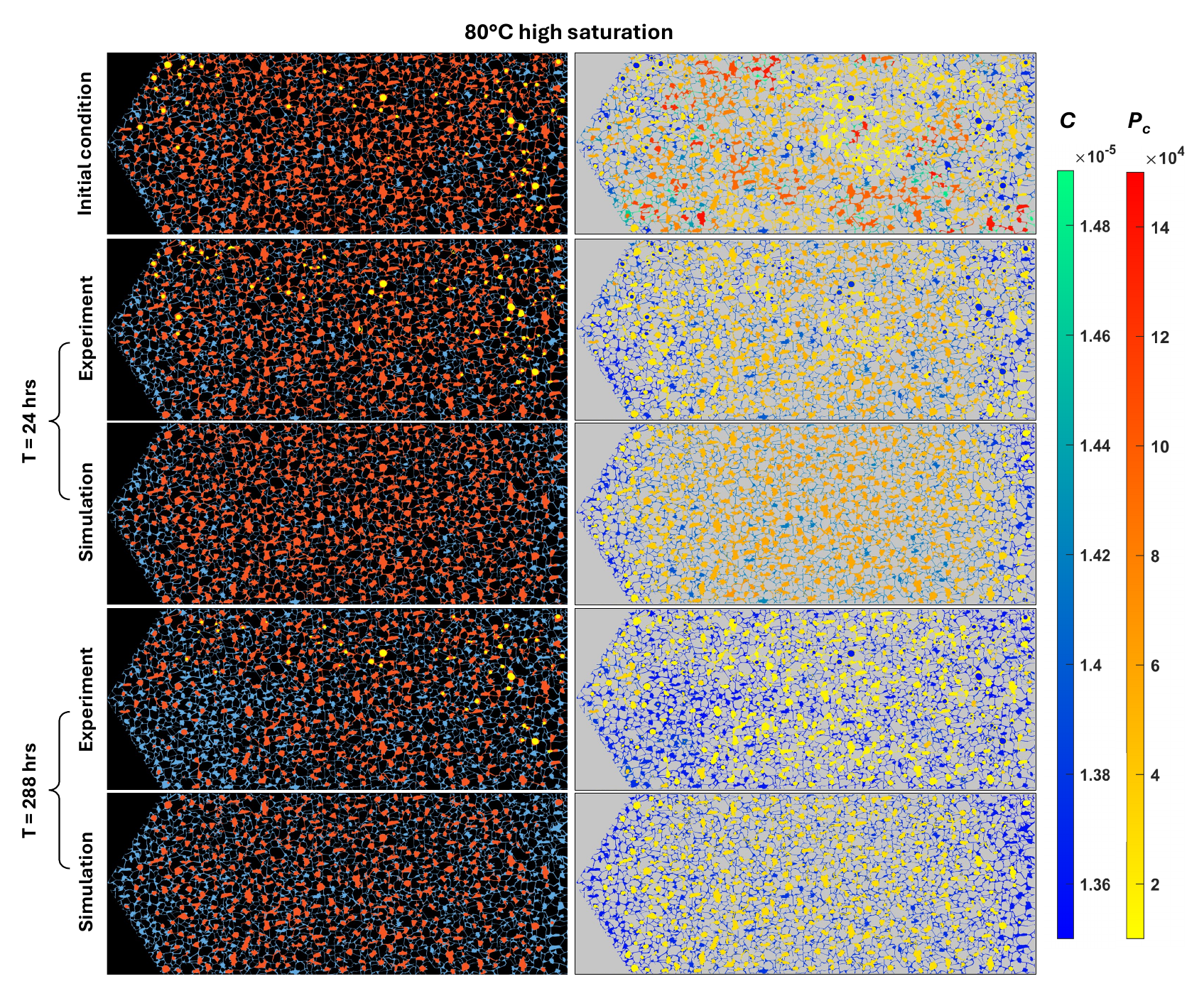}
\caption{Spatial distributions for the 80HS case. Left: phase distribution (red: non-wetting, blue: wetting, black: solid, yellow: condensation). Right: ganglia colored by capillary pressure (dyn/cm$^2$) and wetting phase by dissolved concentration (mole fraction). The top row shows the experimental initial condition used to initialize the simulation. Subsequent rows compare experiments and iPNM at $T \!=\! 24$ and $288$~hrs.}
\label{fig:spt80HS}
\end{figure}

\medskip
\noindent\textbf{Spatial distribution.}
Figs.\ref{fig:spt40HS}--\ref{fig:spt80HS} compare the spatial distribution of trapped ganglia between experiments and iPNM at $T = 24$ and $288$~hrs for the 40HS and 80HS cases. The experimental initial condition used to initialize the simulations is shown in the top row of each figure. These two cases are selected because they illustrate the range of agreement, from strong in 40HS to weak in 80HS. The remaining two cases are included in the supplementary material. The iPNM results are rendered using the connectivity-corrected visualization described in Section~\ref{sec:ipnm_vis} (Fig.\ref{fig:visual}), which maps simulated pore saturations onto the micromodel's pore-scale image. The continuum model does not resolve individual ganglia and is therefore excluded from this comparison.

In Figs.\ref{fig:spt40HS}--\ref{fig:spt80HS}, the ganglia near the channels dissolve more rapidly due to their proximity to concentration sinks, creating a depletion zone that expands inward over time~\cite{mehmani2024deplete}. After $24$~hrs, the depletion zone is already visible in the experiments, and by $288$~hrs, it has expanded significantly. iPNM captures this dissolution front, though the agreement is stronger in 40HS than in 80HS. In the 80HS case, the left portion of the micromodel depletes faster in the experiment than in iPNM, likely due to a low-curvature interface in the bottom part of the left channel that acts as a stronger sink than the uniform-concentration BC assumed in iPNM (see supplementary videos). The right columns of Figs.\ref{fig:spt40HS}--\ref{fig:spt80HS} compare capillary pressure and dissolved concentration fields.
The latter is obtained by solving the Laplace equation over the wetting phase, with concentrations at ganglia set by Henry's law and used as Dirichlet BCs.
In 40HS, these fields from iPNM are in good agreement with experiments. In 80HS, the discrepancy is more pronounced, consistent with idealized BCs at channels and higher prevalence of condensation droplets discussed above.
Another contributing factor to the more rapid depletion observed in iPNM is the PMM curvature bias, which we discuss next.

\begin{figure}[t!]
\centering
\includegraphics[width=0.75\textwidth]{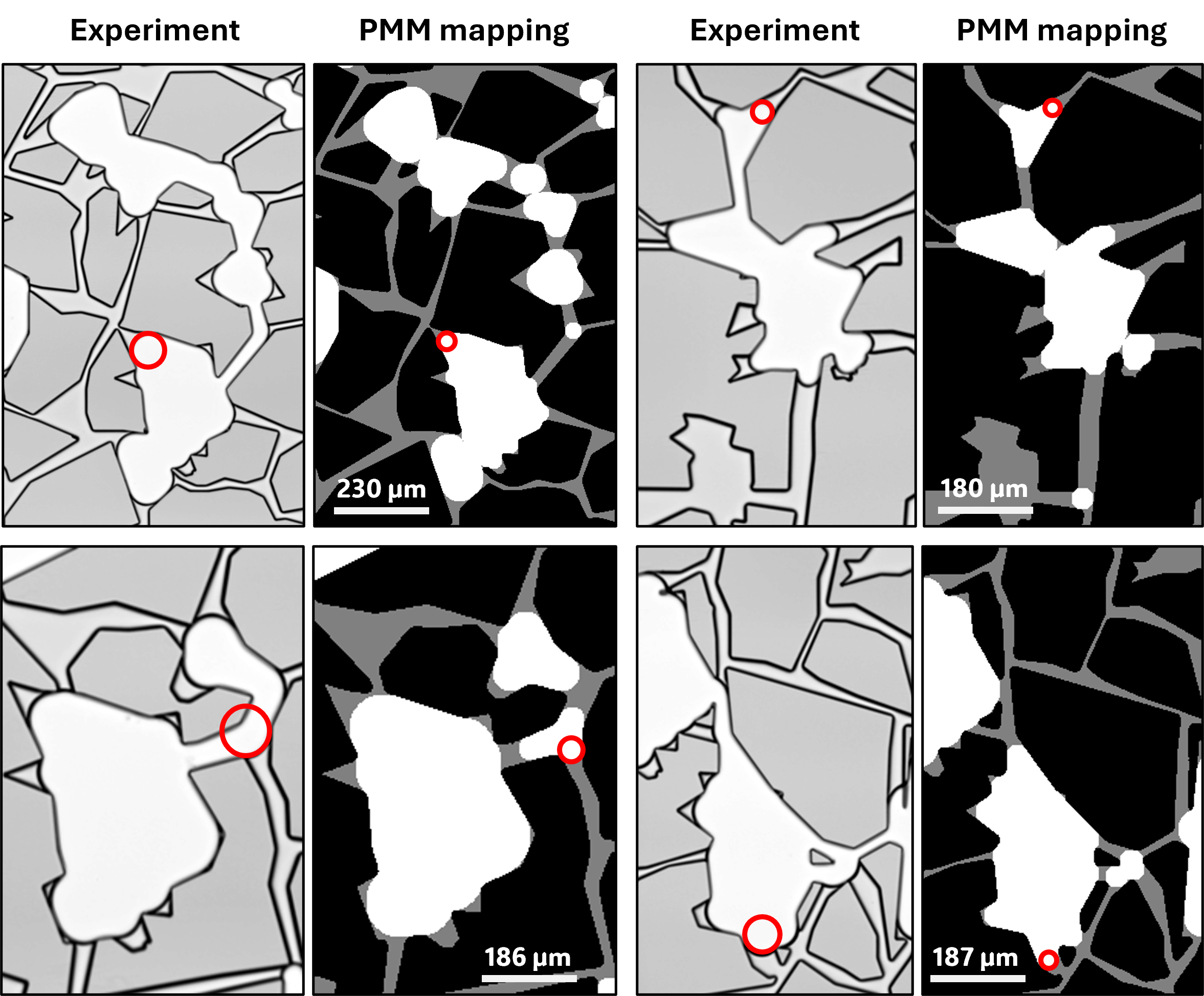}
\caption{Comparison between experimental phase distributions and PMM reconstructions at select pore locations. Red circles are drawn tangent to the gas--water interface. PMM tends to yield smaller radii, indicating a higher curvature than observed experimentally.}
\label{fig:pmmlim}
\end{figure}

\medskip
\noindent\textbf{PMM curvature bias.}
The PMM-derived $\kappa_b$--$S_b$ curves used by iPNM overestimate curvature for pores belonging to multi-pore ganglia, contributing to the underestimation of $\bar{r}$ seen in Figs.\ref{fig:curv}--\ref{fig:curvPDF}. Fig.\ref{fig:pmmlim} demonstrates this visually by comparing experimental phase distributions against PMM reconstructions at the same pore locations. Red circles drawn tangent to various interfaces reveal smaller radii in PMM than in experiments. This arises because PMM constructs the $\kappa_b$--$S_b$ curve for each pore by morphologically eroding its subimage independently. During erosion, connected throats are progressively emptied of the non-wetting phase, so the resulting bubble configuration is confined entirely to the pore body (Fig.\ref{fig:kbsb}). For multi-pore ganglia, however, some of these throats are occupied and hold part of the ganglion's volume. By excluding this throat volume, PMM effectively compresses the same amount of non-wetting phase into a smaller space, producing more deformed interfaces with higher curvature than observed.

\section{Discussion}
\label{sec:disc}

\subsection{Continuum vs. pore-scale modeling}
\label{sec:disc_cont}

We showed in Section~\ref{sec:macro} that the continuum model of~\cite{salehpour2025micro} captures $S_b^{tot}$ well across all four experiments. However, three fundamental limitations distinguish it from iPNM.
First, the continuum model cannot provide pore-scale information about the spatial position, size, curvature, or population statistics of individual ganglia, whereas iPNM resolves all of these by tracking each ganglion explicitly through pore-scale balance equations.
Second, the continuum model assumes capillary equilibrium within an REV at all times, as dictated by the $P_c$--$S_b$ curve. iPNM imposes no such assumption and naturally resolves the pre-equilibration dynamics of ripening. Depending on the size and specific microstructure of the REV, the pre-equilibration timescale can be on the order of macroscopic forcings imposed on the system (e.g., cyclic injections in H$_2$ storage), which would prevent equilibrium from ever being reached. Even more challenging is when an REV itself is absent (e.g., fractal media)~\cite{mehmani2021strive}, for which the continuum framework breaks down entirely.
Third, the $P_c$--$S_b$ curve from PMM, used by the continuum model, corresponds to the \textit{maximal} saturation achievable at any given $P_c$. If the actual saturation is non-maximal, as is often the case due to incomplete sweep during drainage-imbibition cycles, $P_c$ is underestimated~\cite{mehmani2024deplete}. Mitigating this requires extended parameterizations that express $P_c$ as functions of $S_b$, specific interfacial area~\cite{hassan1993pctheory}, and Euler characteristic~\cite{mcclure2018euler,mcclure2020euler}, but these demand evolution equations for these extra variables, which themselves require closure. Pore-scale models like iPNM circumvent this entirely because each ganglion's capillary pressure is computed from its own saturation and pore geometry, with no ambiguity from unoccupied pores. Population balance methods~\cite{bueno2024theory,bueno2025theory} similarly avoid this limitation.

\subsection{iPNM's limitations}
\label{sec:disc_lim}

A key achievement of iPNM is that no geometric approximation in pore shape, typical of all existing PNMs, is introduced. Instead, different pore shapes are encoded by $\kappa_b$--$S_b$ curves obtained from pore-wise applications of PMM. Despite the advance, several limitations remain. One is the PMM curvature bias for multi-pore ganglia identified in Section~\ref{sec:popstat}. Because PMM erodes each pore's subimage independently, connected throats are progressively emptied. This confines the non-wetting phase to the pore body and can overestimate curvature when throats are occupied. One antidote is a connectivity-aware PMM, analogous to the classical invasion-percolation formulation by~\cite{hilpert2001pmm}, but the downside is that the number of occupied-throat permutations grows combinatorially with coordination number. A more tractable approximation is to apply a saturation correction to $\kappa_b$--$S_b$ curves depending on the number of occupied throats. However, the details require further research. Another limitation of iPNM concerns the non-uniqueness of the subcritical branch in the $\kappa_b$--$S_b$ curves for bridge pores (Eq.\ref{eq:kbsb_brg}). As shown in Fig.\ref{fig:kbsb}, multiple interface configurations can exist in a bridge pore depending on which throats are occupied, but all are approximated with the same function. While the correct physical limits are satisfied (Section~\ref{sec:ipnm_pore}), the actual shape between these limits is uncertain.

Other limitations are less consequential for the experiments considered herein but may matter in other settings. First, a zero contact angle was assumed, which is appropriate for the H$_2$--water--silicon system in Section~\ref{sec:exp}. Extensions of PMM to non-zero contact angles exist~\cite{schulz2015pmmCA,liu2022pmmCA} and iPNM can accommodate them without modification. Note the effect enters solely through the $\kappa_b$--$S_b$ curves. Second, while all simulations were on topologically 2D networks, iPNM and its associated PMM operations remain unchanged in 3D. Third, vapor pressure and temperature were assumed spatially uniform, which clearly is insufficient to capture the localized condensation and reevaporation observed at 80\textdegree C. Capturing this phenomenon requires an extra energy balance and the equation of state for water. Lastly, we treated the non-wetting phase density as constant, which was justified by $P_c/P_w \approx 3\%$ in Section~\ref{sec:ipnm_setup}, also typical of the subsurface. When this approximation breaks down, the flow and transport problems become nonlinear.

\subsection{Sources of experimental discrepancy}
\label{sec:disc_exp}

The largest source of discrepancy between iPNM and the experiments is the idealization of the BCs at the channels. In iPNM, the concentration at each channel is set to be uniform and based on the capillary pressure of an interface spanning the channel's half-width (Section~\ref{sec:ipnm_setup}). In experiments, however, the bubble distributions in the channels are highly non-uniform (Fig.\ref{fig:exp}). For example, in the 40HS case, both channels are nearly filled with gas, with the left one containing small condensation droplets that raise the effective capillary pressure above the one assumed in iPNM. In the 80HS case, a low-curvature interface forms in the bottom of the left channel, which acts as a stronger sink than assumed. A similar low-curvature interface sits also in the right channel. These non-uniformities are likely why the micromodel depletes faster in the 80HS experiment than in iPNM (see Fig.\ref{fig:spt80HS} and supplementary videos). The other two cases exhibit similar non-uniformities. In 40LS, the right channel initially contains a small bubble at the bottom that gradually grows upward, lengthening the diffusion path for the topmost ganglia. In 80LS, a low-curvature interface temporarily forms in the right channel but collapses quickly, consistent with the better agreement in $S_b^{tot}$.

Condensation droplets are a secondary source of error, most prominent at 80\textdegree C. In 80HS (Fig.\ref{fig:spt80HS}), droplets repeatedly condense within ganglia and re-evaporate over multiple cycles.
A plausible contributor is small temperature fluctuations across the micromodel. Even on the order of tenths of a degree, such fluctuations can produce sizable swings in vapor pressure via the Clausius-Clapeyron relation ($\sim$4\% per 1\textdegree C).
The elevated gas pressure inside ganglia, by contrast, accounts for only a $\sim$0.05\% shift in vapor pressure due to the Poynting effect.
Moreover, cyclic condensation and re-evaporation perturbs the wetting phase, which can mix dissolved H$_2$ across larger distances. Condensed droplets also increase local ganglion curvature, promoting dissolution by Henry's law. Lastly, the experiments were initialized with deionized water~\cite{salehpour2025micro}, so the wetting phase was undersaturated at $T = 0$, whereas iPNM initializes $C$ from Henry's law assuming local equilibrium. Despite this difference in initialization, no signature is visible in Fig.\ref{fig:curv}, where a temporary drop in mean curvature ($\bar{r}$) would be expected. This effect is therefore of minor importance.

\section{Conclusions}
\label{sec:conclusion}

We presented an image-based pore network model (iPNM) that couples two-phase flow with Ostwald ripening to simulate the evolution of trapped ganglia in heterogeneous porous media. Unlike previous PNMs, iPNM eliminates geometric approximations of pore shape by parameterizing curvature--saturation curves directly from pore-scale images via the pore morphology method (PMM). It handles multi-pore ganglia of arbitrary size and complexity through a combination of pore-type classification, operator-split flow and transport solvers, and an iterative capillary stability checking algorithm that captures invasion, snap-off, retraction, fragmentation, coalescence, and dislocation. The sub-critical branch of the curvature--saturation curve, essential for ripening, is captured for all pore types.

iPNM was verified against a quasi-static PNM to ensure capillary-driven phenomena such as dissolution-induced snap-off, bubble dislocation, and ganglion growth in heterogeneous networks are captured accurately.
iPNM was then validated against microfluidic experiments tracking hydrogen ripening in a sandstone-patterned micromodel over 15--24 days. The model predicts total saturation, mean curvature, curvature distributions, and spatial phase configurations in good agreement with experiments, particularly at 40\textdegree C. At 80\textdegree C, discrepancies arise primarily due to non-ideal boundary conditions at the micromodel's inlet and outlet channels and, to a lesser extent, from condensation droplets not modeled by iPNM. A curvature bias by PMM was also identified for multi-pore ganglia, arising from the independent treatment of each pore during morphological erosion. Comparison with a continuum model confirmed that while macroscopic saturation is captured well by the continuum approach, it cannot resolve population statistics and spatial information about individual ganglia or pre-equilibrium dynamics of ripening within an REV.

Several directions for future work emerge from this work. Extending iPNM to non-zero contact angles, using existing PMM variants, would enable investigating whether ripening dynamics are qualitatively altered under partial wetting. In the context of underground hydrogen storage, generalizing iPNM along the multi-component framework of~\cite{bueno2023PNM} would allow simulating ripening of gas mixtures. Moreover, iPNM opens the possibility of studying how cyclic injection and withdrawal, interacting with ripening, reshape ganglion populations over multiple seasons. Application to 3D rock images extracted from X-ray CT is straightforward, as the PMM operations remain unchanged.

\section*{Supplementary materials} \label{SupMat}
Included as supplementary materials

\section*{Acknowledgments}
This research is jointly supported by the National Science Foundation, United States under Grant No. CBET-2348723 (for ZL, YM) and the Natural Sciences and Engineering Research Council of Canada (NSERC) Alliance Grants under ALLRP 592525-23 and ALLRP 567631-24 (for MS, TL, BZ).
\appendix
\setcounter{figure}{0}
\section{Entry pressure for rectangular throat}
\label{app:entry}

We derive the capillary entry pressure for a throat with rectangular cross-section using the Mayer-Stowe-Princen (MS-P) theory \cite{mayer1965MSP,princen1969MSP,ma1996PceSquare}. The derivation is based on an energy balance at the moment of invasion. Specifically, for a small displacement $\mathrm{d}x$ of the meniscus, the balance of work done versus surface energy created/destroyed reads:
\begin{equation}
\label{eq:MSP_energy}
P_c A_e \,\mathrm{d}x = \left( L_{bw}\sigma_{bw} + L_{bs}\sigma_{bs} - L_{bs}\sigma_{ws} \right) \mathrm{d}x,
\end{equation}
where $A_e$ is the cross-sectional area occupied by the non-wetting phase (effective area), $L_{bw}$ is the length of the bubble-water (non-wetting/wetting) interface, $L_{bs}$ is the length of bubble-solid contact, and $\sigma_{bw}$, $\sigma_{bs}$, $\sigma_{ws}$ are the respective interfacial tensions. Applying Young's equation, $\sigma_{bs} - \sigma_{ws} = \sigma_{bw}\cos\theta$, and denoting $\sigma = \sigma_{bw}$, Eq.\ref{eq:MSP_energy} simplifies to:
\begin{equation}
\label{eq:MSP_simplified}
\frac{P_c}{\sigma} = \frac{L_{bw} + L_{bs}\cos\theta}{A_e} = \frac{P_e}{A_e},
\end{equation}
where $P_e = L_{bw} + L_{bs}\cos\theta$ is the effective perimeter. The next step in the MS-P method is to equate the displacement curvature $P_c/\sigma$ to the curvature $1/r$ of the arc menisci (AMs) residing in the corners \cite{ma1996PceSquare}:
\begin{equation}
\label{eq:MSP_implicit}
\frac{P_c}{\sigma} = \frac{1}{r} = \frac{P_e(r)}{A_e(r)}.
\end{equation}
Eq.\ref{eq:MSP_implicit} is an implicit equation that must be solved for $r$. We next derive explicit expressions for $P_e(r)$ and $A_e(r)$.

Fig.\ref{fig:throat_rect} shows the cross-section of a rectangular throat occupied by the non-wetting phase in the center and wetting phase in the corners, forming four AMs. The throat has an in-plane side length of $2a$ and an out-of-plane side length of $2b$. The distance from the contact line to the corner apex is denoted by $\ell$ and it is related to $r$ via:
\begin{equation}
\label{eq:ell}
\ell = \sqrt{2}\,r\sin\left(\frac{\pi}{4} - \theta\right).
\end{equation}
The bubble-solid contact length $L_{bs}$ is the perimeter of the throat minus the portions occupied by the AMs:
\begin{equation}
\label{eq:Lbs}
L_{bs} = 4\,(a + b - 2\ell).
\end{equation}
The bubble-water arc length $L_{bw}$ counting all four AMs is:
\begin{equation}
\label{eq:Lbw}
L_{bw} = 8\,\left(\frac{\pi}{4} - \theta\right)r.
\end{equation}
Hence, the effective perimeter $P_e$ can be computed via:
\begin{equation}
\label{eq:Pe}
P_e = L_{bw} + L_{bs}\cos\theta = 4\left[2\left(\frac{\pi}{4} - \theta\right)r + (a + b - 2\ell)\cos\theta\right].
\end{equation}

The effective cross-sectional area $A_e$ of the non-wetting phase can be computed by subtracting the four corner-film areas from the throat's cross-sectional area. Straightforward geometric analysis yields the area of a single corner film:
\begin{equation}
\label{eq:Aam}
A_{AM} = \frac{\sqrt{2}}{2}\,\ell \left( \frac{\sqrt{2}}{2}\,\ell + r\cos\left(\frac{\pi}{4} - \theta\right) \right) - \left(\frac{\pi}{4} - \theta\right)r^2.
\end{equation}
which in turns allows computing:
\begin{equation}
\label{eq:Ae}
A_e = 4ab - 4A_{AM} = 4\left[ab - \frac{\sqrt{2}}{2}\,\ell \left( \frac{\sqrt{2}}{2}\,\ell + r\cos\left(\frac{\pi}{4} - \theta\right) \right) + \left(\frac{\pi}{4} - \theta\right)r^2\right].
\end{equation}
Eqs.\ref{eq:Pe} and \ref{eq:Ae} are substituted into Eq.\ref{eq:MSP_implicit} to allow solving for $r$ numerically. The entry pressure is $P_{ce} = \sigma/r$.

\medskip
\noindent\textbf{Special case of square throat.}
For a square throat with $a = b = R$, we show our formulation for rectangular throats reduces to the analytical solution of Ma et al. \cite{ma1996PceSquare}:
\begin{equation}
\label{eq:Pce_square}
\frac{P_{ce}R}{\sigma} = \cos\theta + \sqrt{\,\sin\theta\cos\theta + \frac{\pi}{4} - \theta}.
\end{equation}
Let $\gamma = \pi/4 - \theta$. Substituting $a = b = R$ and $\ell = \sqrt{2}\,r\sin\gamma$ into Eqs.\ref{eq:Pe} and \ref{eq:Ae} yields:
\begin{align}
P_e &= 4\left[2\gamma r + (2R - 2\sqrt{2}\,r\sin\gamma)\cos\theta\right], \\
A_e &= 4\left[R^2 + r^2\gamma - r^2\sin^2\gamma - r^2\sin\gamma\cos\gamma\right].
\end{align}
Inserting into Eq.\ref{eq:MSP_implicit} and using $\cos\theta = \frac{1}{\sqrt{2}}(\cos\gamma + \sin\gamma)$, we obtain after simplification:
\begin{equation} \label{eq:square_quad}
R^2 = r^2[\gamma - \sin\gamma\,(\cos\gamma + \sin\gamma)] + \sqrt{2}\,Rr\,(\cos\gamma + \sin\gamma).
\end{equation}
Defining $u = r/R$, Eq.\ref{eq:square_quad} is a quadratic $Au^2 + Bu - 1 = 0$, with $A = \gamma - \sin\gamma(\cos\gamma + \sin\gamma)$ and $B = \sqrt{2}(\cos\gamma + \sin\gamma)$. The physical root gives:
\begin{equation}
\frac{P_{ce}R}{\sigma} = \frac{1}{u} = \frac{B + \sqrt{B^2 + 4A}}{2}.
\end{equation}
Noting that $B/2 = \cos\theta$ and $B^2 + 4A = 4\sin\theta\cos\theta + 4\,(\pi/4 - \theta)$, this simplifies to Eq.\ref{eq:Pce_square}.

\section{Derivation of non-wetting phase conductivity}
\label{app:cond}

We derive the hydraulic conductance of the non-wetting (or bubble) phase flowing through the central region of an invaded throat with rectangular cross-section (see Fig.\ref{fig:throat_rect}). The derivation employs the hydraulic diameter concept \cite{white2006book} and ensures that the correct asymptotic behavior is preserved as the bubble saturation approaches unity.

To establish some key relations, consider first the scenario where a throat is occupied by one fluid phase undergoing steady-state, fully-developed laminar flow. The corresponding axial momentum balance on a fluid element reads:
\begin{equation}
\label{eq:forcebal}
-\frac{dP}{dx} = \frac{{\tau}_w P_t}{A_t},
\end{equation}
where ${\tau}_w$ is the mean shear stress exerted by the fluid on the throat's walls, $A_t$ is the cross-sectional area of the throat, and $P_t$ is its wetted perimeter. Next, introduce the Fanning friction factor defined as:
\begin{equation}
\label{eq:Cf}
C_f = \frac{\tau_w}{\frac{1}{2}\rho u^2},
\end{equation}
where $\rho$ is the fluid density and $u$ the mean velocity. For laminar flow, $C_f$ scales inversely with Reynolds number:
\begin{equation}
\label{eq:CfRe}
C_f = \frac{Po}{Re}, \quad \text{where} \quad Re = \frac{\rho u D_h}{\mu},
\end{equation}
where $D_h = 4\,A_t/P_t$ is the hydraulic diameter. The Poiseuille number $Po$ is a geometry-dependent constant that characterizes flow resistance for a given throat shape. For example, $Po \!=\! 16$ for a circular duct and $Po \!\approx\! 14.2$ for a square duct \cite{white2006book}.
Substituting Eqs.\ref{eq:Cf}--\ref{eq:CfRe} into Eq.\ref{eq:forcebal} yields:
\begin{equation}
\label{eq:ubar}
u = \frac{D_h^2}{2\,Po\,\mu}\frac{\Delta P}{L_t}.
\end{equation}
The volumetric flow rate can be computed as follows:
\begin{equation}
\label{eq:Qapp}
q_t = u\,A_t = \frac{D_h^2\,A_t}{2\,Po\,\mu}\frac{\Delta P}{L_t}.
\end{equation}
Defining the hydraulic conductance via $q_t = (K_t/\mu)\,\Delta P$, we obtain:
\begin{equation}
\label{eq:Kgen}
K_t = \frac{D_h^2\,A_t}{2\,Po\,L_t}.
\end{equation}
For a rectangular throat with half-widths $a$ and $b$ (Fig.\ref{fig:throat_rect}), $A_t = 4\,ab$, $P_t = 4\,(a+b)$, and $D_h = 4\,A_t/P_t$.

Now for the non-wetting phase in Fig.\ref{fig:throat_rect}, we define the hydraulic conductance in a similar fashion to Eq.\ref{eq:Kgen}:
\begin{equation}
\label{eq:Kbapp}
K_b = \frac{D_{h,b}^2\,A_b}{2\,Po_b\,L_t},
\end{equation}
where $A_b$ is the cross-sectional area of the bubble-occupied region in the throat, $D_{h,b}$ is its hydraulic diameter, and $Po_b$ is its Poiseuille number. This region is bounded by solid walls and the four arc menisci occupying the corners, as shown in Fig.\ref{fig:throat_rect}. Under the assumption of perfect slip at the interface between the bubble and wetting phase, the wetted perimeter $C_b$ consists only of the solid walls. Expressions for $A_b$, $C_b$, and $D_{h,b}$ are given in Section~\ref{sec:ipnm_throat}.

The only quantity on the right-hand side of Eq.\ref{eq:Kbapp} that is not directly available is the Poiseuille number $Po_b$. In general, $Po_b$ depends on the shape of the bubble-occupied region (Fig.\ref{fig:throat_rect}), which varies with the throat's saturation as the corner films grow or shrink. Since no analytical or numerical solution for this dependence exists, we approximate $Po_b$ by the single-phase Poiseuille number $Po$. The latter can be back-calculated using:
\begin{equation}
\label{eq:Poapp}
Po_b \approx Po = \frac{D_h^2\,A_t}{2\,K_t\,L_t}
\end{equation}
where $K_t$ is the known single-phase conductivity from Eq.\ref{eq:Kt}.
This approximation ensures the correct asymptotic behavior: as $S_b \to 1$, the wetting films shrink, causing $A_b \to A_t$, $C_b \to P_t$, and $D_{h,b} \to D_h$, so that $K_b \to K_t$.

\section{Continuum model}
\label{app:continuum}

The continuum model of Salehpour et al.~\cite{salehpour2025micro} describes the macroscopic evolution of the non-wetting phase due to Ostwald ripening. It is derived by taking the continuum limit of the pore-scale mass balance equations of Bueno et al.~\cite{bueno2023PNM}, assuming ideal gas behavior for the non-wetting phase. The full derivation is in the supplementary material of~\cite{salehpour2025micro}. The model is formulated in terms of the dissolved mole fraction $C$ in the aqueous phase. For the micromodel considered here, the domain is treated as 1D in the left-to-right direction $x$ (Fig.~\ref{fig:exp}), and the equation reads:
\begin{equation}
\left\{\left[\frac{H}{\rho_w} - 1\right]\left[H\frac{dS_b}{dP_c} C + S_b\right] + 1\right\}\frac{\partial C}{\partial t} = \frac{D_m}{\tau}\frac{\partial^2 C}{\partial x^2},
\label{eq:continuum}
\end{equation}
where $D_m$ is the molecular diffusion coefficient, $H$ is Henry's constant, $\rho_w$ is the molar density of the wetting phase, and $\tau = \phi^{-1}$ is Bruggeman's tortuosity. The $dS_b/dP_c$ term is obtained from a macroscopic $P_c$--$S_b$ relationship computed by applying PMM to the binary image of the entire micromodel~\cite{mehmani2024deplete}. Unlike classical PMM~\cite{hazlett1995pmm,hilpert2001pmm}, the variant used here retains disconnected ganglia, and unlike the pore-wise application in iPNM (Section~\ref{sec:ipnm_pore}), here PMM is applied globally to produce a single $P_c$--$S_b$ curve. The reader is referred to~\cite{salehpour2025micro} for an illustration of this curve.

The non-wetting phase saturation $S_b$ is recovered from $C$ via Henry's law:
\begin{equation}
C = \frac{P_w + P_c(S_b) - P_v}{H},
\label{eq:henry_cont}
\end{equation}
where $P_w$ is the wetting-phase pressure and $P_v$ is the spatially uniform vapor pressure. To invert Eq.~\ref{eq:henry_cont} for $S_b$, the $P_c$--$S_b$ curve is used. Dirichlet BCs on $C$ are imposed at the left and right ends of the domain by evaluating Eq.~\ref{eq:henry_cont} at $P_c = \sigma(1/R_{l,r} + 2/g)$ from the Young--Laplace equation, where $R_l = 50$~\textmu m and $R_r = 85$~\textmu m are half-widths of the left and right channels, respectively (Fig.~\ref{fig:exp}).
The initial condition for $C$ is obtained by converting the initial non-wetting phase saturation profile to concentration via Eq.~\ref{eq:henry_cont} and the $P_c$--$S_b$ curve. The initial saturation profile itself is constructed from the first experimental image by averaging the phase occupancy in the $y$-direction (up-down in Fig.~\ref{fig:exp}) at each $x$-coordinate using a sliding window of width equal to 5\% of the micromodel length. The total non-wetting phase saturation in the micromodel at each time step is computed by spatially averaging the saturation profile.

Eq.~\ref{eq:continuum} is discretized along the $x$ direction with a uniform mesh of 100 cells. Backward Euler time integration with adaptive time stepping and a Newton solver are used at each time step. Fluid properties are assigned depending on the temperature corresponding to the particular experiment being modeled. These are listed in Table~\ref{tab:fluid}.

\biboptions{numbers,sort&compress}
\bibliographystyle{elsarticle-num}
\bibliography{./References}

\end{document}